\documentclass[aps,pre,preprint,showkeys,nofootinbib]{revtex4}

\usepackage{amsmath,amssymb,amsfonts,graphicx}
\usepackage{array}

\begin{document}

\vspace*{-1cm}

\title{Harmonic Standing-Wave Excitations of Simply-Supported Isotropic Solid Elastic Circular Cylinders: \\ Exact 3D Linear Elastodynamic Response}

\author{Jamal Sakhr}
\author{Blaine A. Chronik}
\affiliation{Department of Physics and Astronomy, The University of Western Ontario, London, Ontario, Canada N6A 3K7}

\date{\today}

\begin{abstract}
The vibration of a solid elastic cylinder is one of the classical applied problems of elastodynamics. Many fundamental forced-vibration problems involving solid elastic cylinders have not yet been studied or solved using the full three-dimensional (3D) theory of linear elasticity. One such problem is the steady-state forced-vibration response of a simply-supported isotropic solid elastic circular cylinder subjected to two-dimensional harmonic standing-wave excitations on its curved surface. In this paper, we exploit certain previously-obtained particular solutions to the Navier-Lam\'{e} equation of motion and exact matrix algebra to construct an exact closed-form 3D elastodynamic solution to the problem. The method of solution is direct and demonstrates a general approach that can be applied to solve other similar forced-vibration problems involving elastic cylinders. Two complete analytical solutions are in fact constructed corresponding to two different but closely-related families of harmonic standing-wave excitations. The second of these analytical solutions is evaluated numerically in order to study the steady-state frequency response in some example excitation cases. In each case, the solution generates a series of resonances that are in correspondence with a subset of the natural frequencies of the simply-supported cylinder. The considered problem is of general interest both as an exactly-solvable 3D elastodynamics problem and as a benchmark forced-vibration problem involving a solid elastic cylinder. 
\end{abstract}

\keywords{Analytically-solvable 3D elastodynamics problems; Benchmark forced-vibration problems; Linear-elastic boundary-value problems with simply-supported boundary conditions; Isotropic solid elastic cylinders; harmonic standing-wave boundary stresses} 

\maketitle

\section{Introduction}
 
The vibration of a solid elastic cylinder is one of the classical applied problems of elastodynamics (see Ref.~\cite{Elastobuch} and references therein) and a problem of fundamental interest to the study of sound and vibration \cite{LeissaQatu11}. 
The literature on the \emph{free} vibration of finite-length solid circular cylinders is vast with many new contributions over the last few decades advocating various methodological approaches to obtaining natural frequencies and mode shapes under a variety of end conditions. Key theoretical contributions to this topic include 
Refs.~\cite{Hutch67,McMahon70,Rumer71,Hutch72,Glad72,Hutch80,Leissa95,Liew95,Loy99,Buchanan01,Kari02,Kari03,Zhou03,Nieves08,Khalili12,Ye14} for isotropic cylinders,  
Refs.~\cite{Lusher88,Chau94,Chen98,Honarvar09} for transversely isotropic cylinders, and Refs.~\cite{Heyliger91,Heyliger92} for anisotropic and inhomogeneous cylinders. (See also Refs.~\cite{Soldatos94,Qatu02} for detailed reviews of work done prior to 2000.) 

There is, in comparison, a scant literature on the \emph{forced} vibrational characteristics of solid elastic cylinders (finite-length, isotropic, circular, or otherwise). The most notable work on this subject is due to Ebenezer and co-workers \cite{Ebenezer05,Ebenezer08}, who devised an exact series method to determine the steady-state forced-vibration response of an isotropic solid elastic cylinder subjected to arbitrary \emph{axisymmetric} excitations on its surfaces. The work of Pan and Pan \cite{Pan98} is also noteworthy since it is a rare example of a study that provides an exact analytical solution to a forced-vibration problem involving elastic cylinders. In Ref.~\cite{Pan98}, Pan and Pan obtained analytically the exact forced response in fixed-free isotropic solid cylinders subjected to purely \emph{torsional} moment excitations. 

As pointed out in Ref.~\cite{VTCS10} and elsewhere, many basic forced-vibration problems involving elastic cylinders have not yet been studied or solved using the full three-dimensional (3D) theory of linear elasticity. One such problem that has hitherto not been considered in the literature is the steady-state vibration response of a simply-supported isotropic solid elastic circular cylinder subjected to two-dimensional (2D) harmonic standing-wave excitations on its curved surface. The posited problem is closely related to the problem of determining the wave-propagation characteristics of \emph{free} harmonic standing waves in a simply-supported isotropic solid elastic circular cylinder; the former is the forced analog of the latter. Explicit formulation and solution of the latter problem is not easy to find in the literature. 
It is however a well-known and often-cited fact that the characteristics of free harmonic standing waves in a finite-length simply-supported elastic cylinder are formally equivalent to those of free harmonic traveling waves in a corresponding cylinder of infinite length, the latter of which are well-studied. Numerical solutions to the \emph{free} standing-wave problem are thus readily available as a corollary \cite{Arm69}. Certain analytical solutions will also be applicable, with certain modifications, in certain special cases (e.g., the Pochhammer-Chree solution will be applicable, with certain modifications, to axisymmetric vibrations) \cite{Soldatos94}. A solution to the analogous \emph{forced}-vibration problem does not however appear to be in the literature. 

Given the mathematical form of the excitations (c.f., Eqs.~(\ref{BCsCRVD1}) and (\ref{BCsCRVD2})), one might expect that the proposed problem can be solved exactly and that the solution has a  closed form. This is indeed the case, and in this paper, we shall exploit certain previously-obtained particular solutions to the Navier-Lam\'{e} equation \cite{usB1} and exact matrix algebra to construct an exact closed-form 3D elastodynamic solution. The method of solution is direct and demonstrates a general approach that can be applied to solve other similar forced-vibration problems involving elastic cylinders. The radial part of the solution, which involves Bessel functions of the first kind, can be expressed in several different but equivalent forms. We make use of certain well-known Bessel function identities in order to cast the radial part of the solution in a symmetric form that is free of derivatives thereby making the obtained analytical solution apt for numerical computation. The analytical solution is later evaluated numerically in order to study the steady-state frequency response of the system in some example cases. In each case, consistency with published natural frequency data is observed. 

The proposed problem is of general interest both as an exactly-solvable 3D elastodynamics problem and as a benchmark forced-vibration problem involving a solid elastic cylinder. The obtained analytical solution is not only useful for revealing the main physical features of the present problem, but can also serve as a benchmark solution for assessment or validation of numerical methods and/or computational software. 

\section{Mathematical Definition of the Problem}\label{EGFULL}

Consider the problem of a simply-supported isotropic solid elastic circular cylinder of finite (but arbitrary) length $L$ and radius $R$ subjected to time-harmonic stresses on its curved surface. These stresses are spatially non-uniform and are such that the circumferential and longitudinal variations are also harmonic. In this paper, we shall work in the circular cylindrical coordinate system wherein all physical quantities depend on the spatial coordinates $(r,\theta,z)$, which denote the radial, circumferential, and longitudinal coordinates, respectively, and on the time $t$. Using this notation, the boundary stresses on the curved surface are: 
\begin{subequations}\label{BCsCRVD1}
\begin{equation}\label{strssrr1}
\sigma_{rr}(R,\theta,z,t)=\mathcal{A}\sin(m\theta)\sin\left({k\pi\over L}z\right)\sin(\omega t),
\end{equation}
\begin{equation}\label{strssrt1}
\sigma_{r\theta}(R,\theta,z,t)=\mathcal{B}\cos(m\theta)\sin\left({k\pi\over L}z\right)\sin(\omega t),
\end{equation}
\begin{equation}\label{strssrz1}
\sigma_{rz}(R,\theta,z,t)=\mathcal{C}\sin(m\theta)\cos\left({k\pi\over L}z\right)\sin(\omega t),
\end{equation}
\end{subequations}
or, 
\begin{subequations}\label{BCsCRVD2}
\begin{equation}\label{strssrr2}
\sigma_{rr}(R,\theta,z,t)=\mathcal{A}\cos(m\theta)\sin\left({k\pi\over L}z\right)\sin(\omega t),
\end{equation}
\begin{equation}\label{strssrt2}
\sigma_{r\theta}(R,\theta,z,t)=\mathcal{B}\sin(m\theta)\sin\left({k\pi\over L}z\right)\sin(\omega t),
\end{equation}
\begin{equation}\label{strssrz2}
\sigma_{rz}(R,\theta,z,t)=\mathcal{C}\cos(m\theta)\cos\left({k\pi\over L}z\right)\sin(\omega t), 
\end{equation}
\end{subequations}
where $\sigma_{rr}(r,\theta,z,t)$ is a normal component of stress, $\sigma_{r\theta}(r,\theta,z,t)$ and $\sigma_{rz}(r,\theta,z,t)$ are shear components of stress, $\{\mathcal{A},\mathcal{B},\mathcal{C}\}$ are prescribed constant stresses, each having units of pressure, $k\in\mathbb{Z}^+$ and $m\in\mathbb{Z}^+\cup\{0\}$ are prescribed dimensionless constants, and $\omega$ is the prescribed angular frequency of excitation. Since the harmonic temporal variation is separable from the harmonic spatial variations in each of the excitations (\ref{strssrr1})-(\ref{strssrz1}) and (\ref{strssrr2})-(\ref{strssrz2}), these represent harmonic standing-wave excitations. 

The specific problem of interest here is to determine the elastodynamic response of the cylinder when it is subjected to the non-uniform distribution of stress (\ref{BCsCRVD1}) or (\ref{BCsCRVD2}) on its curved surface. We thus seek to determine the displacement at all points of the cylinder. The governing equation of motion for the displacement is the Navier-Lam\'{e} (NL) equation, which can be written in vector form as \cite{Elastobuch}:
\begin{equation}\label{NLE}
(\lambda+2\mu)\nabla(\nabla\cdot\mathbf{u})-\mu\nabla\times(\nabla\times\mathbf{u})+\mathbf{b}=\rho{\partial^2\mathbf{u}\over\partial t^2},
\end{equation}
where $\mathbf{u}\equiv\mathbf{u}(r,\theta,z,t)$ is the displacement field, $\lambda>0$ and $\mu>0$ are the first and second Lam\'{e} constants, respectively\footnote{Note that the first Lam\'{e} constant $\lambda$ need not be positive, but we have assumed it to be so for the purposes of this paper.}, and $\rho>0$ is the (constant) density of the cylinder. Since there are only surface forces acting on the cylinder, the local body force is zero (i.e., $\mathbf{b}=0$). The radial, circumferential, and longitudinal components of $\mathbf{u}$ shall here be denoted by $u_r(r,\theta,z,t),u_\theta(r,\theta,z,t)$, and $u_z(r,\theta,z,t)$, respectively. 

Although we have stated that the cylinder is simply supported, we have not yet specified the boundary conditions at the flat ends of the cylinder, which are situated at $z=0$ and $z=L$. The classical simply-supported boundary conditions for the stress and displacement at the flat ends of the cylinder are:
\begin{subequations}\label{SSBCs}
\begin{equation}\label{SSBCs1}
u_r(r,\theta,0,t)=u_r(r,\theta,L,t)=0,
\end{equation}
\begin{equation}\label{SSBCs2}
u_\theta(r,\theta,0,t)=u_\theta(r,\theta,L,t)=0,
\end{equation}
\begin{equation}\label{SSBCs3}
\sigma_{zz}(r,\theta,0,t)=\sigma_{zz}(r,\theta,L,t)=0,
\end{equation}
\end{subequations}
where $\sigma_{zz}(r,\theta,z,t)$ is the normal component of stress along the axis of the cylinder.  

Conditions (\ref{BCsCRVD1}) and (\ref{BCsCRVD2}) must be satisfied for all $\theta\in[0,2\pi]$, $z\in(0,L)$, and arbitrary $t$.  Conditions (\ref{SSBCs1})-(\ref{SSBCs3}) must be satisfied for all $r\in[0,R)$, $\theta\in[0,2\pi]$, and arbitrary $t$. The condition that the displacement field is finite everywhere in the cylinder is implicit. Note that we have not given any information about the displacement field and its time derivatives at some initial time $t=t_0$, and thus the problem as defined is not an initial-boundary-value problem. 

In the following, we shall refer to the linear-elastic boundary-value problem (BVP) defined by (\ref{BCsCRVD1}), (\ref{NLE}), and (\ref{SSBCs}) as BVP 1, and the boundary-value problem defined by (\ref{BCsCRVD2}), (\ref{NLE}), and (\ref{SSBCs}) as BVP 2. Note that, for forced motion, \emph{at least} one of $\{\mathcal{A},\mathcal{B},\mathcal{C}\}$ must be non-zero when $m\neq0$. If $m=0$, then $\mathcal{B}\neq0$ is required in boundary conditions (\ref{BCsCRVD1}) and \emph{at least} one of $\{\mathcal{A},\mathcal{C}\}$ is required to be non-zero in boundary conditions (\ref{BCsCRVD2}). 

For future reference, we cite here the stress-displacement relations from the theory of linear elasticity \cite{VTCS10}:
\begin{subequations}\label{strssdispCYL}
\begin{equation}\label{stssstrnCYL1}
\sigma_{rr}=(\lambda+2\mu){\partial u_r\over\partial r}+{\lambda\over r}\left({\partial u_\theta\over\partial \theta}+u_r\right)+\lambda{\partial u_z\over\partial z},
\end{equation}
\begin{equation}\label{stssstrnCYL2}
\sigma_{\theta\theta}=\lambda{\partial u_r\over\partial r}+{(\lambda+2\mu)\over r}\left({\partial u_\theta\over\partial \theta}+u_r\right)+\lambda{\partial u_z\over\partial z},
\end{equation}
\begin{equation}\label{stssstrnCYL3}
\sigma_{zz}=\lambda{\partial u_r\over\partial r}+{\lambda\over r}\left({\partial u_\theta\over\partial \theta}+u_r\right)+(\lambda+2\mu){\partial u_z\over\partial z},
\end{equation}
\begin{equation}\label{stssstrnCYL4}
\sigma_{r \theta}=\mu\left({1\over r}{\partial u_r\over\partial \theta}+{\partial u_\theta\over\partial r}-{u_\theta\over r}\right)=\sigma_{\theta r},
\end{equation}
\begin{equation}\label{stssstrnCYL5}
\sigma_{rz}=\mu\left({\partial u_r\over\partial z}+{\partial u_z\over\partial r}\right)=\sigma_{zr},
\end{equation}
\begin{equation}\label{stssstrnCYL6}
\sigma_{\theta z}=\mu\left({\partial u_\theta\over\partial z}+{1\over r}{\partial u_z\over\partial \theta}\right)=\sigma_{z\theta}.
\end{equation}
\end{subequations}
Relations (\ref{strssdispCYL}), which provide the general mathematical connection between the components of the displacement and stress fields, will be used extensively in solving the above-defined BVPs.  

Since the excitations [(\ref{BCsCRVD1}) or (\ref{BCsCRVD2})] are time-harmonic, relations (\ref{strssdispCYL}) and solution uniqueness together imply that the response of the cylinder must necessarily be so as well. In other words, the displacement field $\mathbf{u}(r,\theta,z,t)=\mathbf{u}_s(r,\theta,z)\sin(\omega t)$, where $\mathbf{u}_s(r,\theta,z)$ denotes the stationary or time-independent part of the displacement field. It is the latter object that we ultimately seek to determine and to then study.

\section{Some Parametric Solutions to the Navier-Lam\'{e} Equation}\label{SOLNSgen} 

In the absence of body forces, the following parametric solutions to Eq.~(\ref{NLE}) can be obtained using a Buchwald decomposition of the displacement field (see Ref.~\cite{usB1} for details): 
\begin{eqnarray}\label{solnradcomp}
u_r&=&\left(\sum_{s=1}^2\left[a_s\left \{ \begin{array}{c}
             J_n'(\alpha_sr) \\
             I_n'(\alpha_sr)
           \end{array} \right\} + b_s \left \{ \begin{array}{c}
             Y_n'(\alpha_sr) \\
             K_n'(\alpha_sr)
           \end{array} \right\}\right] \Big[c_s\cos(n\theta)+d_s\sin(n\theta)\Big]\right)\phi_z(z)\phi_t(t) \nonumber \\ \nonumber \\
           &+&{n\over r}\left[a_3\left \{ \begin{array}{c}
             J_n(\alpha_2r) \\
             I_n(\alpha_2r)
           \end{array} \right\} + b_3 \left \{ \begin{array}{c}
             Y_n(\alpha_2r) \\
             K_n(\alpha_2r)
           \end{array} \right\}\right] \Big[-c_3\sin(n\theta)+d_3\cos(n\theta)\Big]\chi^{}_z(z)\chi^{}_t(t),   
\end{eqnarray}

\begin{eqnarray}\label{solnangcomp}
u_\theta&=&{n\over r}\left(\sum_{s=1}^2\left[a_s\left \{ \begin{array}{c}
             J_n(\alpha_sr) \\
             I_n(\alpha_sr)
           \end{array} \right\} + b_s \left \{ \begin{array}{c}
             Y_n(\alpha_sr) \\
             K_n(\alpha_sr)
           \end{array} \right\}\right] \Big[-c_s\sin(n\theta)+d_s\cos(n\theta)\Big]\right)\phi_z(z)\phi_t(t) \nonumber \\ \nonumber \\
           &-&\left[a_3\left \{ \begin{array}{c}
             J_n'(\alpha_2r) \\
             I_n'(\alpha_2r)
           \end{array} \right\} + b_3 \left \{ \begin{array}{c}
             Y_n'(\alpha_2r) \\
             K_n'(\alpha_2r)
           \end{array} \right\}\right] \Big[c_3\cos(n\theta)+d_3\sin(n\theta)\Big]\chi^{}_z(z)\chi^{}_t(t),   
\end{eqnarray}
and
\begin{eqnarray}\label{solnaxcomp}
u_z=\left(\sum_{s=1}^2~\gamma_s\left[a_s\left \{ \begin{array}{c}
             J_n(\alpha_sr) \\
             I_n(\alpha_sr)
           \end{array} \right\} + b_s \left \{ \begin{array}{c}
             Y_n(\alpha_sr) \\
             K_n(\alpha_sr)
           \end{array} \right\}\right] \Big[c_s\cos(n\theta)+d_s\sin(n\theta)\Big]\right){\text{d}\psi_z(z)\over\text{d}z}~\psi_t(t), \nonumber \\
\end{eqnarray}
where $n$ is a non-negative integer and $\{a_1,a_2,a_3,b_1,b_2,b_3,c_1,c_2,c_3,d_1,d_2,d_3\}$ are arbitrary constants. The constituents of Eqs.~(\ref{solnradcomp})-(\ref{solnaxcomp}) are as follows:

\noindent \textbf{(i)} The constants $\alpha_1$ and $\alpha_2$ in the arguments of the Bessel functions are given by 
\begin{equation}\label{alphas}
\alpha_1=\sqrt{~\left|\kappa-{\rho\tau\over(\lambda+2\mu)}\right|~}, \quad \alpha_2=\sqrt{~\left|\kappa-{\rho\tau\over\mu}\right|~},
\end{equation}
where $\displaystyle \kappa\in\mathbb{R}\backslash\{0\}$ and $\displaystyle \tau\in\mathbb{R}\backslash\{0\}$ are free parameters. 

\noindent \textbf{(ii)} The correct linear combination of Bessel functions is determined by the relative values of the parameters $\{\lambda,\mu,\rho,\kappa,\tau\}$ as given in Table \ref{TabLinCombos}. 

\vspace*{0.25cm}

\setlength{\extrarowheight}{10pt}
\begin{table}[h]
\centering
\begin{tabular}{| c | c | c |}
    \hline
    ~~~Linear Combination~~~ & ~~~~~ $s=1$ term ~~~~~ & ~~~~~$s=2$ term ~~~~~\\ [10pt] \hline
    $\displaystyle\{J_n(\alpha_sr), Y_n(\alpha_sr)\}$ & $\displaystyle \kappa > {\rho\tau\over(\lambda+2\mu)}$ & $\displaystyle \kappa > {\rho\tau\over\mu}$ \\ [10pt] \hline
    $\displaystyle\{I_n(\alpha_sr), K_n(\alpha_sr)\}$ & $\displaystyle \kappa < {\rho\tau\over(\lambda+2\mu)}$ & $\displaystyle \kappa < {\rho\tau\over\mu}$ \\ [10pt] \hline
\end{tabular}
\caption{Conditions on the radial part of each term in Eqs.~(\ref{solnradcomp})-(\ref{solnaxcomp}).}
\label{TabLinCombos}
\end{table}
\setlength{\extrarowheight}{1pt}

\noindent \textbf{(iii)} In Eqs.~(\ref{solnradcomp})-(\ref{solnangcomp}), primes denote differentiation with respect to the radial coordinate $r$.

\noindent \textbf{(iv)} The functions $\phi_z(z)$, $\phi_t(t)$, $\psi_z(z)$, $\psi_t(t)$, $\chi^{}_z(z)$, and $\chi^{}_t(t)$ are given by 
\begin{eqnarray}\label{Zpart}
\phi_z(z)=\psi_z(z)=\left \{ \begin{array}{lr}
             E\cos\left(\sqrt{|\kappa|}z\right)+F\sin\left(\sqrt{|\kappa|}z\right) & \text{if}~\kappa<0 \\
             E\exp\left(-\sqrt{\kappa}z\right)+F\exp\left(\sqrt{\kappa}z\right) & \text{if}~\kappa>0
           \end{array} \right.,
\end{eqnarray}

\begin{eqnarray}\label{Tpart}
\phi_t(t)=\psi_t(t)=\left \{ \begin{array}{lr}
             G\cos\left(\sqrt{|\tau|}t\right)+H\sin\left(\sqrt{|\tau|}t\right) & \text{if}~\tau<0 \\
             G\exp\left(-\sqrt{\tau}t\right)+H\exp\left(\sqrt{\tau}t\right) & \text{if}~\tau>0
           \end{array} \right.,
\end{eqnarray}

\begin{eqnarray}\label{ZpartC}
\chi^{}_z(z)=\left \{ \begin{array}{lr}
             \widetilde{E}\cos\left(\sqrt{|\kappa|}z\right)+\widetilde{F}\sin\left(\sqrt{|\kappa|}z\right) & \text{if}~\kappa<0 \\
             \widetilde{E}\exp\left(-\sqrt{\kappa}z\right)+\widetilde{F}\exp\left(\sqrt{\kappa}z\right) & \text{if}~\kappa>0
           \end{array} \right.,
\end{eqnarray}

\begin{eqnarray}\label{TpartC}
\chi^{}_t(t)=\left \{ \begin{array}{lr}
             \widetilde{G}\cos\left(\sqrt{|\tau|}t\right)+\widetilde{H}\sin\left(\sqrt{|\tau|}t\right) & \text{if}~\tau<0 \\
             \widetilde{G}\exp\left(-\sqrt{\tau}t\right)+\widetilde{H}\exp\left(\sqrt{\tau}t\right) & \text{if}~\tau>0
           \end{array} \right.,
\end{eqnarray}
where $\displaystyle\left\{E,F,G,H,\widetilde{E},\widetilde{F},\widetilde{G},\widetilde{H}\right\}$ are arbitrary constants. 

\noindent \textbf{(v)} The constant $\gamma_s$ in Eq.~(\ref{solnaxcomp}) is given by 
\begin{eqnarray}\label{solpsiPP2}
\gamma_s=\left \{ \begin{array}{lr}
             1 & \text{if}~s=1 \\
             {1\over\kappa}\left(\kappa - {\rho\tau\over\mu}\right) & \text{if}~s=2
           \end{array} \right.. 
\end{eqnarray}

Note that Eqs.~(\ref{solnradcomp})-(\ref{solnaxcomp}) are valid so long as $(\lambda+2\mu)\kappa \neq {\rho\tau}$ (i.e., $\alpha_1\neq0$) and $\mu\kappa \neq {\rho\tau}$ (i.e., $\alpha_2\neq0$); otherwise the radial parts must be modified as discussed in Ref.~\cite{usB1}. In the following, these conditions will be satisfied, by construction. 

\section{General Form of the Displacement Field}\label{GenFormDispField}

General solutions suited to the boundary-value problems defined in Sec.~\ref{EGFULL} can be easily constructed from the family of parametric solutions given in Sec.~\ref{SOLNSgen} by identifying one or a combination of the physical parameters $\{L,R,k,\omega\}$ with the (free) mathematical parameters $\kappa$ and $\tau$. 
Let $\kappa=-\left({k\pi\over L}\right)^2$ and $\tau=-\omega^2$, and then choose particular solutions defined by taking $E=0$ in Eq.~(\ref{Zpart}), $\widetilde{E}=0$ in Eq.~(\ref{ZpartC}), $G=0$ in Eq.~(\ref{Tpart}), $\widetilde{G}=0$ in Eq.~(\ref{TpartC}), and $n=m$ in Eqs.~(\ref{solnradcomp})-(\ref{solnaxcomp}). Noting the forms of Eqs.~(\ref{solnradcomp})-(\ref{solnaxcomp}) and comparing (\ref{strssrr1})/(\ref{strssrr2}) with (\ref{stssstrnCYL1}), (\ref{strssrt1})/(\ref{strssrt2}) with (\ref{stssstrnCYL4}), and (\ref{strssrz1})/(\ref{strssrz2}) with (\ref{stssstrnCYL5}), we may immediately deduce that the axial and temporal parts of the displacement components are given by 
\begin{equation}\label{phipsiZEG}
\phi_z(z)=\psi_z(z)=F\sin\left({k\pi\over L}z\right), \quad \chi^{}_z(z)=\widetilde{F}\sin\left({k\pi\over L}z\right), 
\end{equation}
\begin{equation}\label{phipsiTEG}
\phi_t(t)=\psi_t(t)=H\sin(\omega t), \quad \chi^{}_t(t)=\widetilde{H}\sin(\omega t).
\end{equation}
By defining a new set of arbitrary constants
\begin{subequations}\label{arbconsts}
\begin{equation}\label{consts1}
\bar{A}_s\equiv a_sc_sFH, \quad \bar{B}_s\equiv b_sc_sFH, \quad (s=1,2)
\end{equation}
\begin{equation}\label{consts2}
\widetilde{A}_s\equiv a_sd_sFH, \quad \widetilde{B}_s\equiv b_sd_sFH, \quad (s=1,2)
\end{equation}
\begin{equation}\label{consts3}
\widetilde{A}_3\equiv a_3c_3\widetilde{F}\widetilde{H}, \quad \widetilde{B}_3\equiv b_3c_3\widetilde{F}\widetilde{H},
\end{equation}
\begin{equation}\label{consts4}
\bar{A}_3\equiv a_3d_3\widetilde{F}\widetilde{H}, \quad \bar{B}_3\equiv b_3d_3\widetilde{F}\widetilde{H},
\end{equation}
\end{subequations}
the following two independent particular solutions may be extracted from Eqs.~(\ref{solnradcomp})-(\ref{solnaxcomp}):
\begin{subequations}\label{GenSolA}
\begin{equation}\label{SolA1}
u_r=\Bigg[\left(\sum_{s=1}^2\widetilde{A}_s\Big\{\text{\large\textcircled{\normalsize 1}}\Big\}_s\right)-\widetilde{A}_3{m\over r}\Big\{\text{\large\textcircled{\normalsize 2}}\Big\}_2\Bigg]\sin(m\theta)\sin\left({k\pi\over L}z\right)\sin(\omega t),   
\end{equation}
\begin{equation}\label{SolA2}
u_\theta=\Bigg[{m\over r}\left(\sum_{s=1}^2\widetilde{A}_s\Big\{\text{\large\textcircled{\normalsize 2}}\Big\}_s\right)-\widetilde{A}_3\Big\{\text{\large\textcircled{\normalsize 1}}\Big\}_2\Bigg]\cos(m\theta)\sin\left({k\pi\over L}z\right)\sin(\omega t),   
\end{equation}
\begin{equation}\label{SolA3}
u_z=\left({k\pi\over L}\right)\Bigg(\sum_{s=1}^2\widetilde{A}_s\gamma_s\Big\{\text{\large\textcircled{\normalsize 2}}\Big\}_s\Bigg)\sin(m\theta)\cos\left({k\pi\over L}z\right)\sin(\omega t),   
\end{equation}
\end{subequations}
and
\begin{subequations}\label{GenSolB}
\begin{equation}\label{SolB1}
u_r=\Bigg[\left(\sum_{s=1}^2\bar{A}_s\Big\{\text{\large\textcircled{\normalsize 1}}\Big\}_s\right)+\bar{A}_3{m\over r}\Big\{\text{\large\textcircled{\normalsize 2}}\Big\}_2\Bigg]\cos(m\theta)\sin\left({k\pi\over L}z\right)\sin(\omega t),   
\end{equation}
\begin{equation}\label{SolB2}
u_\theta=-\Bigg[{m\over r}\left(\sum_{s=1}^2\bar{A}_s\Big\{\text{\large\textcircled{\normalsize 2}}\Big\}_s\right)+\bar{A}_3\Big\{\text{\large\textcircled{\normalsize 1}}\Big\}_2\Bigg]\sin(m\theta)\sin\left({k\pi\over L}z\right)\sin(\omega t),   
\end{equation}
\begin{equation}\label{SolB3}
u_z=\left({k\pi\over L}\right)\Bigg(\sum_{s=1}^2\bar{A}_s\gamma_s\Big\{\text{\large\textcircled{\normalsize 2}}\Big\}_s\Bigg)\cos(m\theta)\cos\left({k\pi\over L}z\right)\sin(\omega t),   
\end{equation}
\end{subequations}
where
\begin{subequations}\label{radpartsGEN}
\begin{eqnarray}\label{radparts1}
\Big\{\text{\large\textcircled{\normalsize 1}}\Big\}_s=\left \{ \begin{array}{c}
             J_m'(\alpha_sr)={m\over r}J_m(\alpha_sr)-\alpha_sJ_{m+1}(\alpha_sr) \\
             I_m'(\alpha_sr)={m\over r}I_m(\alpha_sr)+\alpha_sI_{m+1}(\alpha_sr)
           \end{array} \right\}, \quad \Big\{\text{\large\textcircled{\normalsize 2}}\Big\}_s=\left \{ \begin{array}{c}
             J_m(\alpha_sr) \\
             I_m(\alpha_sr) 
           \end{array} \right\}, ~~~
\end{eqnarray}
\begin{eqnarray}\label{radparts2}
\Big\{\text{\large\textcircled{\normalsize 1}}\Big\}_2=\left \{ \begin{array}{c}
             J_m'(\alpha_2r)={m\over r}J_m(\alpha_2r)-\alpha_2J_{m+1}(\alpha_2r) \\
             I_m'(\alpha_2r)={m\over r}I_m(\alpha_2r)+\alpha_2I_{m+1}(\alpha_2r) 
           \end{array} \right\}, \quad \Big\{\text{\large\textcircled{\normalsize 2}}\Big\}_2=\left \{ \begin{array}{c}
             J_m(\alpha_2r) \\
             I_m(\alpha_2r) 
           \end{array} \right\}. ~~~
\end{eqnarray}
\end{subequations}
Since the Bessel functions of the second kind $\{Y_p(\alpha_sr),K_p(\alpha_sr)\}\to\pm\infty$ as $r\to0$ ($p\ge0$), the finiteness condition dictates that all such radial terms should be discarded leaving only Bessel functions of the first kind. 
Note that particular solutions (\ref{GenSolA}) and (\ref{GenSolB}) automatically satisfy boundary conditions (\ref{SSBCs1}) and (\ref{SSBCs2}). Note also that (by virtue of (\ref{stssstrnCYL3})) $\sigma_{zz}(r,\theta,z,t)=F(r,\theta)\sin\left({k\pi\over L}z\right)\sin(\omega t)$, where the precise form of $F(r,\theta)$ is not relevant for our purposes, and thus displacements (\ref{GenSolA}) and (\ref{GenSolB}) as well automatically satisfy boundary condition (\ref{SSBCs3}). 

It can be deduced from inspection of (\ref{strssdispCYL}), (\ref{GenSolA}), and (\ref{GenSolB}), that solution (\ref{GenSolA}) is appropriate for BVP 1, whereas solution (\ref{GenSolB}) is appropriate for BVP 2. The proper choices of Bessel functions in the radial parts of the displacement components depend on the relative values of the material and excitation parameters; three cases can be distinguished as listed in Table \ref{EG1Cases}. The problem should be solved separately for each of these three cases. Note that there are two special cases not included in Table \ref{EG1Cases}: (i) ${\rho\omega^2\over(\lambda+2\mu)}=\left({k\pi\over L}\right)^2$; and (ii) ${\rho\omega^2\over\mu}=\left({k\pi\over L}\right)^2$. These two singular cases require special treatment and shall not be considered here. As a final remark, note that solutions (\ref{GenSolA}) and (\ref{GenSolB}) are valid only when $k\in\mathbb{Z}^+$. 

\setlength{\extrarowheight}{10pt}
\begin{table}[h]
\centering
\begin{tabular}{| c | c | c |} \hline 
    ~Case~ & ~Parametric Relationship ($k$, $\omega$)~ & ~Parametric Relationship ($\kappa$, $\tau$)~ \\ [5pt] \hline 
     1 & $\displaystyle {\rho\omega^2\over(\lambda+2\mu)}<{\rho\omega^2\over\mu}<\left({k\pi\over L}\right)^2$ & $\displaystyle \kappa<{\rho\tau\over(\lambda+2\mu)}$ and $\displaystyle \kappa<{\rho\tau\over\mu}$ \\ [10pt] \hline 
     2 & $\displaystyle \left({k\pi\over L}\right)^2<{\rho\omega^2\over(\lambda+2\mu)}<{\rho\omega^2\over\mu}$ & $\displaystyle \kappa>{\rho\tau\over(\lambda+2\mu)}$ and $\displaystyle \kappa>{\rho\tau\over\mu}$ \\ [10pt] \hline 
     3 & $\displaystyle {\rho\omega^2\over(\lambda+2\mu)}<\left({k\pi\over L}\right)^2<{\rho\omega^2\over\mu}$ & $\displaystyle \kappa<{\rho\tau\over(\lambda+2\mu)}$ and $\displaystyle \kappa>{\rho\tau\over\mu}$ \\ [10pt] \hline 
\end{tabular}
\caption{Parametric relationships defining three distinct sub-problems. In the second column, the relationship is expressed in terms of the physical excitation parameters $k$ and $\omega$, whereas in the third column, the relationship is expressed in terms of the mathematical parameters $\kappa$ and $\tau$.}
\label{EG1Cases}
\end{table} 
\setlength{\extrarowheight}{1pt}

\section{Analytical Solution to BVP 1}\label{solntoBVP1}

\subsection{\textbf{Case 1:} $\displaystyle{\rho\omega^2\over(\lambda+2\mu)}<{\rho\omega^2\over\mu}<\left({k\pi\over L}\right)^2$}\label{BVP1C1}

In this case, $\kappa<{\rho\tau/(\lambda+2\mu)}$ and $\kappa<{\rho\tau/\mu}$ (c.f., Table \ref{EG1Cases}), and thus, according to Table \ref{TabLinCombos}, the modified Bessel functions $I_m(\alpha_sr)$ (and their derivatives) should be employed in the radial parts of Eqs.~(\ref{GenSolA}), where the constants $\alpha_1$ and $\alpha_2$, as determined from Eq.~(\ref{alphas}), are:
\begin{equation}\label{alphasEG}
\alpha_1=\sqrt{\left({k\pi\over L}\right)^2-{\rho\omega^2\over(\lambda+2\mu)}}, \quad \alpha_2=\sqrt{\left({k\pi\over L}\right)^2-{\rho\omega^2\over\mu}}.
\end{equation}
The constant $\gamma_s$ in Eq.~(\ref{SolA3}), as determined from Eq.~(\ref{solpsiPP2}), is given by 
\begin{eqnarray}\label{solpsiPP2EG}
\displaystyle
\gamma_s=\left \{ \begin{array}{lr}
             1 & ~~\text{if}~s=1 \\
             1-\left[{\rho\omega^2\over\left({k\pi\over L}\right)^2\mu}\right] & ~~\text{if}~s=2
           \end{array} \right..
\end{eqnarray}
Inputting the above ingredients into (\ref{GenSolA}), the displacement components take the form:
\begin{subequations}\label{SolAC1}
\begin{equation}\label{SolAC11}
u_r=\Bigg\{\sum_{s=1}^2\widetilde{A}_s\bigg[{m\over r}I_m(\alpha_sr)+\alpha_sI_{m+1}(\alpha_sr)\bigg]-\widetilde{A}_3{m\over r}I_m(\alpha_2r)\Bigg\}\sin(m\theta)\sin\left({k\pi\over L}z\right)\sin(\omega t), 
\end{equation}
\vspace*{-0.1cm}
\begin{equation}\label{SolAC12}
u_\theta=\Bigg\{{m\over r}\left(\sum_{s=1}^2\widetilde{A}_sI_m(\alpha_sr)\right)-\widetilde{A}_3\bigg[{m\over r}I_m(\alpha_2r)+\alpha_2I_{m+1}(\alpha_2r)\bigg]\Bigg\}\cos(m\theta)\sin\left({k\pi\over L}z\right)\sin(\omega t),  
\end{equation}
\vspace*{-0.1cm}
\begin{equation}\label{SolAC13}
u_z=\left({k\pi\over L}\right)\Bigg\{\sum_{s=1}^2\widetilde{A}_s\gamma_sI_m(\alpha_sr)\Bigg\}\sin(m\theta)\cos\left({k\pi\over L}z\right)\sin(\omega t),   
\end{equation}
\end{subequations}
where the constants $\alpha_s$ and $\gamma_s$ are given by Eqs.~(\ref{alphasEG}) and (\ref{solpsiPP2EG}), respectively. 

To complete the solution, we must determine the values of the constants $\left\{\widetilde{A}_1,\widetilde{A}_2,\widetilde{A}_3\right\}$ in Eqs.~(\ref{SolAC11})-(\ref{SolAC13}) that satisfy boundary conditions (\ref{BCsCRVD1}). Substituting Eqs.~(\ref{SolAC11})-(\ref{SolAC13}) into Eqs.~(\ref{stssstrnCYL1}), (\ref{stssstrnCYL4}), and (\ref{stssstrnCYL5}), and performing the lengthy algebra yields the stress components:
\begin{subequations}\label{strsscmpRRC1}
\begin{equation}\label{strsscmpRRC11A}
\setlength{\jot}{5pt}
\begin{split}
\sigma_{rr}(r,\theta,z,t)=\Bigg\{\sum_{s=1}^2\widetilde{A}_s\left[\left(\beta_s+2\mu{m(m-1)\over r^2}\right)I_m(\alpha_sr)-{2\mu\alpha_s\over r}I_{m+1}(\alpha_sr)\right]  \\ 
- 2\mu \widetilde{A}_3\left[{m(m-1)\over r^2}I_m(\alpha_2r)+{\alpha_2m\over r}I_{m+1}(\alpha_2r)\right]\Bigg\}\sin(m\theta)\sin\left({k\pi\over L}z\right)\sin(\omega t),
\end{split}
\end{equation}
where
\begin{equation}\label{strsscmpRRC11B}
\beta_s=\lambda\left[\alpha^2_s-\gamma_s\left({k\pi\over L}\right)^2\right]+2\mu\alpha^2_s, \quad s=1,2
\end{equation}
\end{subequations} 
\begin{equation}\label{strsscmpRTC1}
\setlength{\jot}{5pt}
\begin{split}
\sigma_{r\theta}(r,\theta,z,t)=2\mu\Bigg\{\sum_{s=1}^2\widetilde{A}_s\bigg[{m(m-1)\over r^2}I_m(\alpha_sr)+{\alpha_sm\over r}I_{m+1}(\alpha_sr)\bigg] \quad \quad \\ 
+\widetilde{A}_3\left[-\bigg({\alpha_2^2\over2}+{m(m-1)\over r^2}\bigg)I_m(\alpha_2r)+{\alpha_2\over r}I_{m+1}(\alpha_2r)\right]\Bigg\}\cos(m\theta)\sin\left({k\pi\over L}z\right)\sin(\omega t),
\end{split}
\end{equation}
and 
\begin{equation}\label{strsscmpRZC1}
\setlength{\jot}{5pt}
\begin{split}
\sigma_{rz}(r,\theta,z,t)=\mu\left({k\pi\over L}\right)\Bigg\{\sum_{s=1}^2\widetilde{A}_s\left(1+\gamma_s\right)\bigg({m\over r}I_m(\alpha_sr)+\alpha_sI_{m+1}(\alpha_sr)\bigg) \\
-\widetilde{A}_3\bigg({m\over r}I_m(\alpha_2r)\bigg)\Bigg\}\sin(m\theta)\cos\left({k\pi\over L}z\right)\sin(\omega t).
\end{split} 
\end{equation}

When $m\neq0$, application of the boundary conditions proceeds by substituting Eqs.~(\ref{strsscmpRRC1}), (\ref{strsscmpRTC1}), and (\ref{strsscmpRZC1}) into the LHSs of Eqs.~(\ref{strssrr1}), (\ref{strssrt1}), and (\ref{strssrz1}), respectively, and then canceling identical sinusoidal terms on both sides of the resulting equations. This yields the following conditions: 
\begin{subequations}\label{BCsC1app}
\begin{equation}\label{BCsC1app1}
\setlength{\jot}{5pt}
\begin{split}
\sum_{s=1}^2\widetilde{A}_s\left[\left(\beta_s+2\mu{m(m-1)\over R^2}\right)I_m(\alpha_sR)-{2\mu\alpha_s\over R}I_{m+1}(\alpha_sR)\right]  \\ 
- 2\mu \widetilde{A}_3\left[{m(m-1)\over R^2}I_m(\alpha_2R)+{\alpha_2m\over R}I_{m+1}(\alpha_2R)\right]=\mathcal{A},
\end{split}
\end{equation}
\begin{equation}\label{BCsC1app2}
\setlength{\jot}{5pt}
\begin{split}
\sum_{s=1}^2\widetilde{A}_s\bigg[{m(m-1)\over R^2}I_m(\alpha_sR)+{\alpha_sm\over R}I_{m+1}(\alpha_sR)\bigg] \quad \quad \\ 
+\widetilde{A}_3\left[-\bigg({\alpha_2^2\over2}+{m(m-1)\over R^2}\bigg)I_m(\alpha_2R)+{\alpha_2\over R}I_{m+1}(\alpha_2R)\right]={\mathcal{B}\over2\mu},
\end{split}
\end{equation} 
\begin{equation}\label{BCsC1app3}
\sum_{s=1}^2\widetilde{A}_s\left(1+\gamma_s\right)\bigg({m\over R}I_m(\alpha_sR)+\alpha_sI_{m+1}(\alpha_sR)\bigg) 
-\widetilde{A}_3\bigg({m\over R}I_m(\alpha_2R)\bigg)={\mathcal{C}\over\mu\left({k\pi\over L}\right)}.
\end{equation}
\end{subequations} 
Conditions (\ref{BCsC1app}) can be written as the $3\times3$ linear system:
\begin{subequations}\label{BCmatEQC1}
\begin{eqnarray}\label{BCmatEQC11}
\left[\begin{array}{ccc}
            f_m-v_{m+1} & g_m-w_{m+1} & -\Big((m-1)q_m+mw_{m+1}\Big) \\
            (m-1)p_m+mv_{m+1} & ~(m-1)q_m+mw_{m+1}~ & -\Big(h_m-w_{m+1}\Big) \\
           2\big(p_m+v_{m+1}\big) & (1+\gamma_2)\big(q_m+w_{m+1}\big) & -q_m
           \end{array} \right] \left[\begin{array}{c}
            \widetilde{A}_1 \\
            \widetilde{A}_2 \\
            \widetilde{A}_3
           \end{array} \right] =  \left[\begin{array}{c}
           \mathbb{A} \\
           \mathbb{B} \\
           \mathbb{C}    
           \end{array} \right], \nonumber \\
\end{eqnarray}
where
\begin{equation}\label{BCmatEQC12a}
f_m\equiv\left[{\beta_1R\over2\mu}+{m(m-1)\over R}\right]I_m(\alpha_1R), 
\end{equation}
\begin{equation}\label{BCmatEQC12b}
g_m\equiv\left[{\beta_2R\over2\mu}+{m(m-1)\over R}\right]I_m(\alpha_2R), 
\end{equation}
\begin{equation}\label{BCmatEQC12c}
h_m\equiv\left[{\alpha^2_2R\over2}+{m(m-1)\over R}\right]I_m(\alpha_2R),
\end{equation}
\begin{equation}\label{BCmatEQC13}
p_m\equiv{m\over R}I_m(\alpha_1R), \quad q_m\equiv{m\over R}I_m(\alpha_2R),
\end{equation}
\begin{equation}\label{BCmatEQC14}
v_{m+1}\equiv\alpha_1I_{m+1}(\alpha_1R), \quad w_{m+1}\equiv\alpha_2I_{m+1}(\alpha_2R),
\end{equation}
\begin{equation}\label{BCmatEQC15}
\mathbb{A}\equiv\left({\mathcal{A}\over2\mu}\right)R, \quad \mathbb{B}\equiv\left({\mathcal{B}\over2\mu}\right)R, \quad \mathbb{C}\equiv\left({\mathcal{C}\over k\pi\mu}\right)L.
\end{equation}
\end{subequations}

\vspace*{0.15cm}

\noindent The general solution to system (\ref{BCmatEQC1}) can be expressed in the form:
\begin{subequations}\label{solnBVP1C1m}
\begin{equation}\label{solnP1C1m1}
\widetilde{A}_i={\widetilde{\delta}_i\over\mathbb{D}}\Big(\mathsf{C}^{\scriptscriptstyle (i)}_\mathbb{A}\mathbb{A}+\mathsf{C}^{\scriptscriptstyle (i)}_\mathbb{B}\mathbb{B}+\mathsf{C}^{\scriptscriptstyle (i)}_\mathbb{C}\mathbb{C}\Big), \quad i=1,2,3,
\end{equation}
where
\begin{eqnarray}\label{solnP1C1m0}
\widetilde{\delta}_i=\left\{\begin{array}{lll}
             -1 & ~~\text{if}~i=1 \\
             +1 & ~~\text{if}~i=2 \\
             +1 & ~~\text{if}~i=3 \\
           \end{array} \right.,
\end{eqnarray}
\end{subequations}
and the values of the coefficients $\left\{\mathsf{C}^{\scriptscriptstyle (i)}_\mathbb{A},\mathsf{C}^{\scriptscriptstyle (i)}_\mathbb{B},\mathsf{C}^{\scriptscriptstyle (i)}_\mathbb{C}:~i=1,2,3\right\}$ and determinant $\mathbb{D}$ obtained using exact algebra are given by (\ref{solncoeffsC1are}) in Appendix \ref{solncoeffs}. 

\subsubsection{Special Case: $\displaystyle m=0$}

When $m=0$, $\sigma_{rr}=\sigma_{rz}=0$ and $u_r=u_z=0$. Substituting Eq.~(\ref{strsscmpRTC1}) into the LHS of boundary condition (\ref{strssrt1}) and then canceling identical sinusoidal terms on both sides of the resulting equation yields the coefficient $\widetilde{A}_3$ in the remaining displacement component $u_\theta$, which from Eq.~(\ref{SolAC12}) then reduces to
\begin{equation}\label{C1SCm0}
u_\theta(r,\theta,z,t)=\left({(\mathcal{B}/2\mu)\over{\alpha_2\over2}I_0(\alpha_2R)-{1\over R}I_1(\alpha_2R)}\right)I_1(\alpha_2r)\sin\left({k\pi\over L}z\right)\sin(\omega t).
\end{equation}

\subsubsection{Example Case: $\displaystyle m=1$}

When $m=1$, the coefficients $\left\{\mathsf{C}^{\scriptscriptstyle (i)}_\mathbb{A},\mathsf{C}^{\scriptscriptstyle (i)}_\mathbb{B},\mathsf{C}^{\scriptscriptstyle (i)}_\mathbb{C}:~i=1,2,3\right\}$ reduce to: 
\begin{subequations}\label{solnBVP1C1meq1} 
\begin{equation}\label{solnP2C1meq12}
\mathsf{C}^{\scriptscriptstyle (1)}_\mathbb{A}=q_1w_{2}-(1+\gamma_2)\big(h_1-w_{2}\big)\big(q_1+w_{2}\big),
\end{equation}
\begin{equation}\label{solnP2C1meq13}
\mathsf{C}^{\scriptscriptstyle (1)}_\mathbb{B}=(1+\gamma_2)\big(q_1+w_{2}\big)w_{2}-\big(g_1-w_{2}\big)q_1,
\end{equation}
\begin{equation}\label{solnP2C1meq14}
\mathsf{C}^{\scriptscriptstyle (1)}_\mathbb{C}=\big(g_1-w_{2}\big)\big(h_1-w_{2}\big)-w_{2}w_{2},
\end{equation}
\begin{equation}\label{solnP2C1meq15}
\mathsf{C}^{\scriptscriptstyle (2)}_\mathbb{A}=q_1v_{2}-2\big(h_1-w_{2}\big)\big(p_1+v_{2}\big),
\end{equation}
\begin{equation}\label{solnP2C1meq16}
\mathsf{C}^{\scriptscriptstyle (2)}_\mathbb{B}=2\big(p_1+v_{2}\big)w_{2}-\big(f_1-v_{2}\big)q_1,
\end{equation}
\begin{equation}\label{solnP2C1meq17}
\mathsf{C}^{\scriptscriptstyle (2)}_\mathbb{C}=\big(f_1-v_{2}\big)\big(h_1-w_{2}\big)-v_{2}w_{2},
\end{equation}
\begin{equation}\label{solnP2C1meq18}
\mathsf{C}^{\scriptscriptstyle (3)}_\mathbb{A}=(1+\gamma_2)\big(q_1+w_{2}\big)v_{2}-2\big(p_1+v_{2}\big)w_{2},
\end{equation}
\begin{equation}\label{solnP2C1meq19}
\mathsf{C}^{\scriptscriptstyle (3)}_\mathbb{B}=2\big(g_1-w_{2}\big)\big(p_1+v_{2}\big)-(1+\gamma_2)\big(f_1-v_{2}\big)\big(q_1+w_{2}\big),
\end{equation}
\begin{equation}\label{solnP2C1meq110}
\mathsf{C}^{\scriptscriptstyle (3)}_\mathbb{C}=\big(f_1-v_{2}\big)w_{2}-\big(g_1-w_{2}\big)v_{2},
\end{equation}
and the determinant $\mathbb{D}$ reduces to: 
\begin{eqnarray}\label{solnP2C1meq111}
\mathbb{D}&=&2\Big[w_{2}w_{2}-\big(g_1-w_{2}\big)(h_1-w_{2}\big)\Big]\big(p_1+v_{2}\big) \nonumber \\ 
&~+&(1+\gamma_2)\Big[\big(f_1-v_{2}\big)\big(h_1-w_{2}\big)-v_{2}w_{2}\Big]\big(q_1+w_{2}\big) \nonumber \\ 
&~~+&\Big[\big(g_1-w_{2}\big)v_{2}-\big(f_1-v_{2}\big)w_{2}\Big]q_1.
\end{eqnarray}
\end{subequations} 

\subsection{\textbf{Case 2:} $\displaystyle \left({k\pi\over L}\right)^2<{\rho\omega^2\over(\lambda+2\mu)}<{\rho\omega^2\over\mu}$}\label{BVP1C2}

According to Tables \ref{TabLinCombos} and \ref{EG1Cases}, the Bessel functions $J_m(\alpha_sr)$ (and their derivatives) should in this case be employed in the radial parts of Eqs.~(\ref{GenSolA}). The  displacement components thus take the form:
\begin{subequations}\label{GenSolAC2}
\begin{equation}\label{SolAC21}
u_r=\Bigg\{\sum_{s=1}^2\widetilde{A}_s\bigg[{m\over r}J_m(\alpha_sr)-\alpha_sJ_{m+1}(\alpha_sr)\bigg]-\widetilde{A}_3{m\over r}J_m(\alpha_2r)\Bigg\}\sin(m\theta)\sin\left({k\pi\over L}z\right)\sin(\omega t),   
\end{equation}
\vspace*{-0.1cm}
\begin{equation}\label{SolAC22}
u_\theta=\Bigg\{{m\over r}\left(\sum_{s=1}^2\widetilde{A}_sJ_m(\alpha_sr)\right)-\widetilde{A}_3\bigg[{m\over r}J_m(\alpha_2r)-\alpha_2J_{m+1}(\alpha_2r)\bigg]\Bigg\}\cos(m\theta)\sin\left({k\pi\over L}z\right)\sin(\omega t),   
\end{equation}
\vspace*{-0.1cm}
\begin{equation}\label{SolAC23}
u_z=\left({k\pi\over L}\right)\Bigg\{\sum_{s=1}^2\widetilde{A}_s\gamma_sJ_m(\alpha_sr)\Bigg\}\sin(m\theta)\cos\left({k\pi\over L}z\right)\sin(\omega t),   
\end{equation}
\end{subequations}
where 
\begin{equation}\label{alphasEG5}
\alpha_1=\sqrt{-\left({k\pi\over L}\right)^2+{\rho\omega^2\over(\lambda+2\mu)}}, \quad \alpha_2=\sqrt{-\left({k\pi\over L}\right)^2+{\rho\omega^2\over\mu}},
\end{equation}
and $\gamma_s$ is again given by Eq.~(\ref{solpsiPP2EG}). 

The constants $\left\{\widetilde{A}_1,\widetilde{A}_2,\widetilde{A}_3\right\}$ in Eqs.~(\ref{SolAC21})-(\ref{SolAC23}) must as before be chosen so as to satisfy boundary conditions (\ref{BCsCRVD1}). Proceeding as in the previous case, we first obtain general formulas for the radial components of the stress field. Substituting Eqs.~(\ref{SolAC21})-(\ref{SolAC23}) into Eqs.~(\ref{stssstrnCYL1}), (\ref{stssstrnCYL4}), and (\ref{stssstrnCYL5}), and performing the lengthy algebra yields the required stress components:
\begin{subequations}\label{strsscmpRRC2P1}
\begin{equation}\label{strsscmpRRC2P1A}
\setlength{\jot}{5pt}
\begin{split}
\sigma_{rr}(r,\theta,z,t)=\Bigg\{\sum_{s=1}^2\widetilde{A}_s\left[\left(-\eta_s+2\mu{m(m-1)\over r^2}\right)J_m(\alpha_sr)+{2\mu\alpha_s\over r}J_{m+1}(\alpha_sr)\right]  \\ 
+ 2\mu \widetilde{A}_3\left[-{m(m-1)\over r^2}J_m(\alpha_2r)+{\alpha_2m\over r}J_{m+1}(\alpha_2r)\right]\Bigg\}\sin(m\theta)\sin\left({k\pi\over L}z\right)\sin(\omega t),
\end{split}
\end{equation}
where
\begin{equation}\label{strsscmpRRC2P1B}
\eta_s=\lambda\left[\alpha^2_s+\gamma_s\left({k\pi\over L}\right)^2\right]+2\mu\alpha^2_s, \quad s=1,2
\end{equation}
\end{subequations} 
\begin{equation}\label{strsscmpRTC2P1}
\setlength{\jot}{5pt}
\begin{split}
\sigma_{r\theta}(r,\theta,z,t)=2\mu\Bigg\{\sum_{s=1}^2\widetilde{A}_s\bigg[{m(m-1)\over r^2}J_m(\alpha_sr)-{\alpha_sm\over r}J_{m+1}(\alpha_sr)\bigg] \quad \quad \\ 
+\widetilde{A}_3\left[\bigg({\alpha_2^2\over2}-{m(m-1)\over r^2}\bigg)J_m(\alpha_2r)-{\alpha_2\over r}J_{m+1}(\alpha_2r)\right]\Bigg\}\cos(m\theta)\sin\left({k\pi\over L}z\right)\sin(\omega t),
\end{split}
\end{equation}
and 
\begin{equation}\label{strsscmpRZC2P1}
\setlength{\jot}{5pt}
\begin{split}
\sigma_{rz}(r,\theta,z,t)=\mu\left({k\pi\over L}\right)\Bigg\{\sum_{s=1}^2\widetilde{A}_s\left(1+\gamma_s\right)\bigg({m\over r}J_m(\alpha_sr)-\alpha_sJ_{m+1}(\alpha_sr)\bigg) \\
-\widetilde{A}_3\bigg({m\over r}J_m(\alpha_2r)\bigg)\Bigg\}\sin(m\theta)\cos\left({k\pi\over L}z\right)\sin(\omega t).
\end{split} 
\end{equation}

When $m\neq0$, application of the boundary conditions (\ref{BCsCRVD1}) as described in Sec.~\ref{BVP1C1} yields three conditions involving the constants $\left\{\widetilde{A}_1,\widetilde{A}_2,\widetilde{A}_3\right\}$. Collectively, these three conditions can be written as the $3\times3$ linear system:
\begin{subequations}\label{BCmatEQP1C2}
\begin{eqnarray}\label{BCmatEQP1C21}
\left[\begin{array}{ccc}
            F_m+V_{m+1} & G_m+W_{m+1} & -\Big((m-1)Q_m-mW_{m+1}\Big) \\
            (m-1)P_m-mV_{m+1} & ~(m-1)Q_m-mW_{m+1}~ & -\Big(H_m+W_{m+1}\Big) \\
           2\big(P_m-V_{m+1}\big) & (1+\gamma_2)\big(Q_m-W_{m+1}\big) & -Q_m
           \end{array} \right] \left[\begin{array}{c}
            \widetilde{A}_1 \\
            \widetilde{A}_2 \\
            \widetilde{A}_3
           \end{array} \right] =  \left[\begin{array}{c}
           \mathbb{A} \\
           \mathbb{B} \\
           \mathbb{C}    
           \end{array} \right], \nonumber \\
\end{eqnarray}
where
\begin{equation}\label{BCmatEQP1C2a}
F_m\equiv\left[-\left({\eta_1R\over2\mu}\right)+{m(m-1)\over R}\right]J_m(\alpha_1R), 
\end{equation}
\begin{equation}\label{BCmatEQP1C2b}
G_m\equiv\left[-\left({\eta_2R\over2\mu}\right)+{m(m-1)\over R}\right]J_m(\alpha_2R), 
\end{equation}
\begin{equation}\label{BCmatEQP1C2c}
H_m\equiv\left[-\left({\alpha^2_2R\over2}\right)+{m(m-1)\over R}\right]J_m(\alpha_2R),
\end{equation}
\begin{equation}\label{BCmatEQP1C23}
P_m\equiv{m\over R}J_m(\alpha_1R), \quad Q_m\equiv{m\over R}J_m(\alpha_2R),
\end{equation}
\begin{equation}\label{BCmatEQP1C24}
V_{m+1}\equiv\alpha_1J_{m+1}(\alpha_1R), \quad W_{m+1}\equiv\alpha_2J_{m+1}(\alpha_2R),
\end{equation}
\end{subequations}
and $\{\mathbb{A},\mathbb{B},\mathbb{C}\}$ are as given by (\ref{BCmatEQC15}). 

The general solution to system (\ref{BCmatEQP1C2}) can again be expressed in the form given by Eq.~(\ref{solnBVP1C1m}) with the exact values of the coefficients $\left\{\mathsf{C}^{\scriptscriptstyle (i)}_\mathbb{A},\mathsf{C}^{\scriptscriptstyle (i)}_\mathbb{B},\mathsf{C}^{\scriptscriptstyle (i)}_\mathbb{C}:~i=1,2,3\right\}$ and determinant $\mathbb{D}$ in this case given by (\ref{solncoeffsC2are}) in Appendix \ref{solncoeffs}.

\subsubsection{Special Case: $\displaystyle m=0$}

When $m=0$, $\sigma_{rr}=\sigma_{rz}=0$ and $u_r=u_z=0$. Using Eq.~(\ref{strsscmpRTC2P1}) in boundary condition (\ref{strssrt1}) yields the coefficient $\widetilde{A}_3$ in the remaining displacement component $u_\theta$, which from Eq.~(\ref{SolAC22}) then reduces to
\begin{equation}\label{C2SCm0}
u_\theta(r,\theta,z,t)=\left({(\mathcal{B}/2\mu)\over{\alpha_2\over2}J_0(\alpha_2R)-{1\over R}J_1(\alpha_2R)}\right)J_1(\alpha_2r)\sin\left({k\pi\over L}z\right)\sin(\omega t).
\end{equation}

\subsubsection{Example Case: $\displaystyle m=1$}

When $m=1$, the coefficients $\left\{\mathsf{C}^{\scriptscriptstyle (i)}_\mathbb{A},\mathsf{C}^{\scriptscriptstyle (i)}_\mathbb{B},\mathsf{C}^{\scriptscriptstyle (i)}_\mathbb{C}:~i=1,2,3\right\}$ reduce to: 
\begin{subequations}\label{solnBVP1C2meq1}
\begin{equation}\label{solnP2C2meq12}
\mathsf{C}^{\scriptscriptstyle (1)}_\mathbb{A}=-Q_1W_{2}-(1+\gamma_2)\big(H_1+W_{2}\big)\big(Q_1-W_{2}\big),
\end{equation}
\begin{equation}\label{solnP2C2meq13}
\mathsf{C}^{\scriptscriptstyle (1)}_\mathbb{B}=-(1+\gamma_2)\big(Q_1-W_{2}\big)W_{2}-\big(G_1+W_{2}\big)Q_1,
\end{equation}
\begin{equation}\label{solnP2C2meq14}
\mathsf{C}^{\scriptscriptstyle (1)}_\mathbb{C}=\big(G_1+W_{2}\big)\big(H_1+W_{2}\big)-W_{2}W_{2},
\end{equation}
\begin{equation}\label{solnP2C2meq15}
\mathsf{C}^{\scriptscriptstyle (2)}_\mathbb{A}=-Q_1V_{2}-2\big(H_1+W_{2}\big)\big(P_1-V_{2}\big),
\end{equation}
\begin{equation}\label{solnP2C2meq16}
\mathsf{C}^{\scriptscriptstyle (2)}_\mathbb{B}=-2\big(P_1-V_{2}\big)W_{2}-\big(F_1+V_{2}\big)Q_1,
\end{equation}
\begin{equation}\label{solnP2C2meq17}
\mathsf{C}^{\scriptscriptstyle (2)}_\mathbb{C}=\big(F_1+V_{2}\big)\big(H_1+W_{2}\big)-V_{2}W_{2},
\end{equation}
\begin{equation}\label{solnP2C2meq18}
\mathsf{C}^{\scriptscriptstyle (3)}_\mathbb{A}=-(1+\gamma_2)\big(Q_1-W_{2}\big)V_{2}+2\big(P_1-V_{2}\big)W_{2},
\end{equation}
\begin{equation}\label{solnP2C2meq19}
\mathsf{C}^{\scriptscriptstyle (3)}_\mathbb{B}=2\big(G_1+W_{2}\big)\big(P_1-V_{2}\big)-(1+\gamma_2)\big(F_1+V_{2}\big)\big(Q_1-W_{2}\big),
\end{equation}
\begin{equation}\label{solnP2C2meq110}
\mathsf{C}^{\scriptscriptstyle (3)}_\mathbb{C}=-\big(F_1+V_{2}\big)W_{2}+\big(G_1+W_{2}\big)V_{2},
\end{equation}
and the determinant $\mathbb{D}$ reduces to: 
\begin{eqnarray}\label{sonP2C2meq111}
&&\mathbb{D}=2\Big[W_{2}W_{2}-\big(G_1+W_{2}\big)\big(H_1+W_{2}\big)\Big]\big(P_1-V_{2}\big) \nonumber \\ 
&&+~(1+\gamma_2)\Big[\big(F_1+V_{2}\big)\big(H_1+W_{2}\big)-V_{2}W_{2}\Big]\big(Q_1-W_{2}\big) \nonumber \\ 
&&+~\Big[-\big(G_1+W_{2}\big)V_{2}+\big(F_1+V_{2}\big)W_{2}\Big]Q_1. 
\end{eqnarray}
\end{subequations} 

\subsection{\textbf{Case 3:} $\displaystyle {\rho\omega^2\over(\lambda+2\mu)}<\left({k\pi\over L}\right)^2<{\rho\omega^2\over\mu}$}\label{BVP1C3}

According to Tables \ref{TabLinCombos} and \ref{EG1Cases}, the $s=1$ term in each of the radial parts of Eqs.(\ref{SolA1})-(\ref{SolA3}) should employ the modified Bessel functions $I_m(\alpha_1r)$ (and their derivatives) while the $s=2$ terms should employ the Bessel functions $J_m(\alpha_2r)$ (and their derivatives). The remaining terms are unmodified from those of Case 2. The displacement components thus take the form:
\begin{subequations}\label{GenSolAC3}
\begin{eqnarray}\label{SolAC31}
u_r&=&\Bigg\{\widetilde{A}_1\bigg[{m\over r}I_m(\alpha_1r)+\alpha_1I_{m+1}(\alpha_1r)\bigg]+\widetilde{A}_2\bigg[{m\over r}J_m(\alpha_2r)-\alpha_2J_{m+1}(\alpha_2r)\bigg] \nonumber \\ 
&& \quad -\widetilde{A}_3{m\over r}J_m(\alpha_2r)\Bigg\}\sin(m\theta)\sin\left({k\pi\over L}z\right)\sin(\omega t),   
\end{eqnarray}
\vspace*{-0.1cm}
\begin{eqnarray}\label{SolAC32}
u_\theta&=&\Bigg\{{m\over r}\left(\widetilde{A}_1I_m(\alpha_1r)+\widetilde{A}_2J_m(\alpha_2r)\right)-\widetilde{A}_3\bigg[{m\over r}J_m(\alpha_2r)-\alpha_2J_{m+1}(\alpha_2r)\bigg]\Bigg\} \nonumber \\
&& \quad \quad \quad \quad \times \cos(m\theta)\sin\left({k\pi\over L}z\right)\sin(\omega t), 
\end{eqnarray}
\vspace*{-0.1cm}
\begin{equation}\label{SolAC33}
u_z=\left({k\pi\over L}\right)\Bigg\{\widetilde{A}_1\gamma_1I_m(\alpha_1r)+\widetilde{A}_2\gamma_2J_m(\alpha_2r)\Bigg\}\sin(m\theta)\cos\left({k\pi\over L}z\right)\sin(\omega t),   
\end{equation}
\end{subequations}
where 
\begin{equation}\label{alphasEG3}
\alpha_1=\sqrt{\left({k\pi\over L}\right)^2-{\rho\omega^2\over(\lambda+2\mu)}}, \quad \alpha_2=\sqrt{-\left({k\pi\over L}\right)^2+{\rho\omega^2\over\mu}},
\end{equation}
and $\gamma_s$ is again given by Eq.~(\ref{solpsiPP2EG}).

The constants $\left\{\widetilde{A}_1,\widetilde{A}_2,\widetilde{A}_3\right\}$ in Eqs.~(\ref{SolAC31})-(\ref{SolAC33}) must as before be chosen so as to satisfy boundary conditions (\ref{BCsCRVD1}). Substituting Eqs.~(\ref{SolAC31})-(\ref{SolAC33}) into Eqs.~(\ref{stssstrnCYL1}), (\ref{stssstrnCYL4}), and (\ref{stssstrnCYL5}), and performing the lengthy algebra yields the required stress components:
\begin{subequations}\label{strsscmpRRC3}
\begin{eqnarray}\label{strsscmpRRC3P1}
&&\sigma_{rr}(r,\theta,z,t)=\Bigg\{\widetilde{A}_1\left[\left(\beta_1+2\mu{m(m-1)\over r^2}\right)I_m(\alpha_1r)-{2\mu\alpha_1\over r}I_{m+1}(\alpha_1r)\right] \nonumber \\  
&&+ ~\widetilde{A}_2\left[\left(-\eta_2+2\mu{m(m-1)\over r^2}\right)J_m(\alpha_2r)+{2\mu\alpha_2\over r}J_{m+1}(\alpha_2r)\right] \nonumber \\
&&+ ~2\mu \widetilde{A}_3\left[-{m(m-1)\over r^2}J_m(\alpha_2r)+{\alpha_2m\over r}J_{m+1}(\alpha_2r)\right]\Bigg\}\sin(m\theta)\sin\left({k\pi\over L}z\right)\sin(\omega t), \quad \quad 
\end{eqnarray}
where
\begin{equation}\label{strsscmpRRC3P2}
\beta_1=\lambda\left[\alpha^2_1-\gamma_1\left({k\pi\over L}\right)^2\right]+2\mu\alpha^2_1, \quad \eta_2=\lambda\left[\alpha^2_2+\gamma_2\left({k\pi\over L}\right)^2\right]+2\mu\alpha^2_2, 
\end{equation}
\end{subequations} 
\begin{eqnarray}\label{strsscmpRTC3}
&&\sigma_{r\theta}(r,\theta,z,t)=2\mu\Bigg\{\widetilde{A}_1\bigg[{m(m-1)\over r^2}I_m(\alpha_1r)+{\alpha_1m\over r}I_{m+1}(\alpha_1r)\bigg] \nonumber \\
&&+ ~\widetilde{A}_2\bigg[{m(m-1)\over r^2}J_m(\alpha_2r)-{\alpha_2m\over r}J_{m+1}(\alpha_2r)\bigg] \nonumber \\ 
&&+ ~\widetilde{A}_3\left[\bigg({\alpha_2^2\over2}-{m(m-1)\over r^2}\bigg)J_m(\alpha_2r)-{\alpha_2\over r}J_{m+1}(\alpha_2r)\right]\Bigg\}\cos(m\theta)\sin\left({k\pi\over L}z\right)\sin(\omega t), \quad \quad \quad 
\end{eqnarray}
and 
\begin{eqnarray}\label{strsscmpRZC3}
&&\sigma_{rz}(r,\theta,z,t)=\mu\left({k\pi\over L}\right)\Bigg\{\widetilde{A}_1\left(1+\gamma_1\right)\bigg({m\over r}I_m(\alpha_1r)+\alpha_1I_{m+1}(\alpha_1r)\bigg) \nonumber \\ 
&&+ ~\widetilde{A}_2\left(1+\gamma_2\right)\bigg({m\over r}J_m(\alpha_2r)-\alpha_2J_{m+1}(\alpha_2r)\bigg) 
-  \widetilde{A}_3\bigg({m\over r}J_m(\alpha_2r)\bigg)\Bigg\} \sin(m\theta)\cos\left({k\pi\over L}z\right)\sin(\omega t). \nonumber \\ 
\end{eqnarray} 

When $m\neq0$, application of the boundary conditions (\ref{BCsCRVD1}) as described in Sec.~\ref{BVP1C1} yields three conditions, which can be written as the $3\times3$ linear system:
\begin{eqnarray}\label{BCmatEQP1C3}
\left[\begin{array}{ccc}
            f_m-v_{m+1} & G_m+W_{m+1} & -\Big((m-1)Q_m-mW_{m+1}\Big) \\
            (m-1)p_m+mv_{m+1} & ~(m-1)Q_m-mW_{m+1}~ & -\Big(H_m+W_{m+1}\Big) \\
           2\big(p_m+v_{m+1}\big) & (1+\gamma_2)\big(Q_m-W_{m+1}\big) & -Q_m
           \end{array} \right] \left[\begin{array}{c}
            \widetilde{A}_1 \\
            \widetilde{A}_2 \\
            \widetilde{A}_3
           \end{array} \right] =  \left[\begin{array}{c}
           \mathbb{A} \\
           \mathbb{B} \\
           \mathbb{C}    
           \end{array} \right], \nonumber \\
\end{eqnarray}
where all matrix elements have been previously defined by Eqs.~(\ref{BCmatEQC12a})-(\ref{BCmatEQC15}) and Eqs.~(\ref{BCmatEQP1C2a})-(\ref{BCmatEQP1C24}). 

The general solution to system (\ref{BCmatEQP1C3}) can once more be expressed in the form given by Eq.~(\ref{solnBVP1C1m}) with the exact values of the coefficients $\left\{\mathsf{C}^{\scriptscriptstyle (i)}_\mathbb{A},\mathsf{C}^{\scriptscriptstyle (i)}_\mathbb{B},\mathsf{C}^{\scriptscriptstyle (i)}_\mathbb{C}:~i=1,2,3\right\}$ and determinant $\mathbb{D}$ in this case given by (\ref{solncoeffsC3are}) in Appendix \ref{solncoeffs}.

\subsubsection{Special Case: $\displaystyle m=0$}

When $m=0$, $\sigma_{rr}=\sigma_{rz}=0$ and $u_r=u_z=0$. It can be shown that application of boundary condition (\ref{strssrt1}) yields the same result for $u_\theta$ as that found for Case 2, namely, Eq.~(\ref{C2SCm0}). 

\subsubsection{Example Case: $\displaystyle m=1$}

When $m=1$, the coefficients $\left\{\mathsf{C}^{\scriptscriptstyle (i)}_\mathbb{A},\mathsf{C}^{\scriptscriptstyle (i)}_\mathbb{B},\mathsf{C}^{\scriptscriptstyle (i)}_\mathbb{C}:~i=1,2,3\right\}$ reduce to: 
\begin{subequations}\label{solnBVP1C3meq1}
\begin{equation}\label{solnP2C3meq12}
\mathsf{C}^{\scriptscriptstyle (1)}_\mathbb{A}=-Q_1W_{2}-(1+\gamma_2)\big(H_1+W_{2}\big)\big(Q_1-W_{2}\big),
\end{equation}
\begin{equation}\label{solnP2C3meq13}
\mathsf{C}^{\scriptscriptstyle (1)}_\mathbb{B}=-(1+\gamma_2)\big(Q_1-W_{2}\big)W_{2}-\big(G_1+W_{2}\big)Q_1,
\end{equation}
\begin{equation}\label{solnP2C3meq14}
\mathsf{C}^{\scriptscriptstyle (1)}_\mathbb{C}=\big(G_1+W_{2}\big)\big(H_1+W_{2}\big)-W_{2}W_{2},
\end{equation}
\begin{equation}\label{solnP2C3meq15}
\mathsf{C}^{\scriptscriptstyle (2)}_\mathbb{A}=Q_1v_{2}-2\big(H_1+W_{2}\big)\big(p_1+v_{2}\big),
\end{equation}
\begin{equation}\label{solnP2C3meq16}
\mathsf{C}^{\scriptscriptstyle (2)}_\mathbb{B}=-2\big(p_1+v_{2}\big)W_{2}-\big(f_1-v_{2}\big)Q_1,
\end{equation}
\begin{equation}\label{solnP2C3meq17}
\mathsf{C}^{\scriptscriptstyle (2)}_\mathbb{C}=\big(f_1-v_{2}\big)\big(H_1+W_{2}\big)+v_{2}W_{2},
\end{equation}
\begin{equation}\label{solnP2C3meq18}
\mathsf{C}^{\scriptscriptstyle (3)}_\mathbb{A}=(1+\gamma_2)\big(Q_1-W_{2}\big)v_{2}+2\big(p_1+v_{2}\big)W_{2},
\end{equation}
\begin{equation}\label{solnP2C3meq19}
\mathsf{C}^{\scriptscriptstyle (3)}_\mathbb{B}=2\big(G_1+W_{2}\big)\big(p_1+v_{2}\big)-(1+\gamma_2)\big(f_1-v_{2}\big)\big(Q_1-W_{2}\big),
\end{equation}
\begin{equation}\label{solnP2C3meq110}
\mathsf{C}^{\scriptscriptstyle (3)}_\mathbb{C}=-\big(f_1-v_{2}\big)W_{2}-\big(G_1+W_{2}\big)v_{2},
\end{equation}
and the determinant $\mathbb{D}$ reduces to: 
\begin{eqnarray}\label{sonP2C3meq111}
&&\mathbb{D}=2\Big[W_{2}W_{2}-\big(G_1+W_{2}\big)\big(H_1+W_{2}\big)\Big]\big(p_1+v_{2}\big) \nonumber \\ 
&&+~(1+\gamma_2)\Big[\big(f_1-v_{2}\big)\big(H_1+W_{2}\big)+v_{2}W_{2}\Big]\big(Q_1-W_{2}\big) \nonumber \\ 
&&+~\Big[\big(G_1+W_{2}\big)v_{2}+\big(f_1-v_{2}\big)W_{2}\Big]Q_1. 
\end{eqnarray}
\end{subequations} 

\section{Analytical Solution to BVP 2}\label{solntoBVP2}

\subsection{\textbf{Case 1:} $\displaystyle{\rho\omega^2\over(\lambda+2\mu)}<{\rho\omega^2\over\mu}<\left({k\pi\over L}\right)^2$}

Employing (\ref{GenSolB}) and following the logic at the beginning of Sec.~\ref{BVP1C1}, the displacement components take the form:
\begin{subequations}\label{SolBC1}
\begin{equation}\label{SolBC11}
u_r=\Bigg\{\sum_{s=1}^2\bar{A}_s\bigg[{m\over r}I_m(\alpha_sr)+\alpha_sI_{m+1}(\alpha_sr)\bigg]+\bar{A}_3{m\over r}I_m(\alpha_2r)\Bigg\}\cos(m\theta)\sin\left({k\pi\over L}z\right)\sin(\omega t), 
\end{equation}
\vspace*{-0.1cm}
\begin{equation}\label{SolBC12}
u_\theta=-\Bigg\{{m\over r}\left(\sum_{s=1}^2\bar{A}_sI_m(\alpha_sr)\right)+\bar{A}_3\bigg[{m\over r}I_m(\alpha_2r)+\alpha_2I_{m+1}(\alpha_2r)\bigg]\Bigg\}\sin(m\theta)\sin\left({k\pi\over L}z\right)\sin(\omega t),  
\end{equation}
\vspace*{-0.1cm}
\begin{equation}\label{SolBC13}
u_z=\left({k\pi\over L}\right)\Bigg\{\sum_{s=1}^2\bar{A}_s\gamma_sI_m(\alpha_sr)\Bigg\}\cos(m\theta)\cos\left({k\pi\over L}z\right)\sin(\omega t),   
\end{equation}
\end{subequations}
where the constants $\alpha_s$ and $\gamma_s$ are given by Eqs.~(\ref{alphasEG}) and (\ref{solpsiPP2EG}), respectively. 

To determine the values of the constants $\left\{\bar{A}_1,\bar{A}_2,\bar{A}_3\right\}$ in Eqs.~(\ref{SolBC11})-(\ref{SolBC13}) that satisfy boundary conditions (\ref{BCsCRVD2}), we follow the same procedure outlined in Sec.~\ref{BVP1C1}. This leads to the $3\times3$ linear system:
\begin{eqnarray}\label{BC2matEQC1}
\left[\begin{array}{ccc}
            f_m-v_{m+1} & g_m-w_{m+1} & (m-1)q_m+mw_{m+1} \\
            (m-1)p_m+mv_{m+1} & ~(m-1)q_m+mw_{m+1}~ & h_m-w_{m+1} \\
           2\big(p_m+v_{m+1}\big) & (1+\gamma_2)\big(q_m+w_{m+1}\big) & q_m
           \end{array} \right] \left[\begin{array}{c}
            \bar{A}_1 \\
            \bar{A}_2 \\
            \bar{A}_3
           \end{array} \right] =  \left[\begin{array}{r}
           \mathbb{A}~ \\
           -\mathbb{B}~ \\
           \mathbb{C}~    
           \end{array} \right], 
\end{eqnarray}
where the entries in the above $3\times3$ coefficient matrix are given by Eqs.~(\ref{BCmatEQC12a})-(\ref{BCmatEQC14}), and $\{\mathbb{A},\mathbb{B},\mathbb{C}\}$ are given by (\ref{BCmatEQC15}). The general solution to system (\ref{BC2matEQC1}) is:
\begin{subequations}\label{solnBVP2C1m}
\begin{equation}\label{solnP2C1m1}
\bar{A}_i={\bar{\delta}_i\over\mathbb{D}}\Big(\mathsf{C}^{\scriptscriptstyle (i)}_\mathbb{A}\mathbb{A}-\mathsf{C}^{\scriptscriptstyle (i)}_\mathbb{B}\mathbb{B}+\mathsf{C}^{\scriptscriptstyle (i)}_\mathbb{C}\mathbb{C}\Big), \quad i=1,2,3,
\end{equation}
where
\begin{eqnarray}\label{solnP2C1m0}
\bar{\delta}_i=\left\{\begin{array}{ll}
             +1 & ~~\text{even}~i \\
              -1 & ~~\text{odd}~i
           \end{array} \right.,
\end{eqnarray}
\end{subequations} 
the coefficients $\left\{\mathsf{C}^{\scriptscriptstyle (i)}_\mathbb{A},\mathsf{C}^{\scriptscriptstyle (i)}_\mathbb{B},\mathsf{C}^{\scriptscriptstyle (i)}_\mathbb{C}:~i=1,2,3\right\}$ are given by Eqs.~(\ref{solnP2C1m2})-(\ref{solnP2C1m10}), and $\mathbb{D}$ is given by Eq.~(\ref{sonP2C1m11}). 

\subsubsection{Special Case: $\displaystyle m=0$}\label{BVP2axysolnC1}

When $m=0$, $u_\theta=0$, $\sigma_{r\theta}=0$, and boundary condition (\ref{strssrt2}) is identically satisfied. Application of boundary conditions (\ref{strssrr2}) and (\ref{strssrz2}) yields the $2\times2$ linear system:  
\begin{eqnarray}\label{BCmatEQC1SCm0}
\left[\begin{array}{cc}
            f_0-v_1 & g_0-w_1 \\
             2v_1 & ~\big(1+\gamma_2\big)w_1~
           \end{array} \right] \left[\begin{array}{c}
             \bar{A}_1 \\
             \bar{A}_2
           \end{array} \right] =  \left[\begin{array}{c}
              \mathbb{A} \\
              \mathbb{C}
           \end{array} \right].
\end{eqnarray}
The general solution to system (\ref{BCmatEQC1SCm0}) is:
\begin{subequations}\label{BVP2C1meq0}
\begin{equation}\label{BVPC1meq01}
\bar{A}_1={\big(1+\gamma_2\big)w_1\mathbb{A}-\big(g_0-w_1\big)\mathbb{C}\over\big(1+\gamma_2\big)w_1\big(f_0-v_1\big)-2v_1\big(g_0-w_1\big)},
\end{equation}
\begin{equation}\label{BVPC1meq02}
\bar{A}_2={\big(f_0-v_1\big)\mathbb{C}-2v_1\mathbb{A}\over\big(1+\gamma_2\big)w_1\big(f_0-v_1\big)-2v_1\big(g_0-w_1\big)},
\end{equation}
\end{subequations}
where $\left\{f_0,g_0\right\}$ and $\left\{v_1,w_1\right\}$ are the evaluated zero- and first-order Bessel functions (respectively) obtained from substituting $m=0$ in Eqs.~(\ref{BCmatEQC12a}), (\ref{BCmatEQC12b}), and (\ref{BCmatEQC14}). 

Thus, in this special case, the non-zero components of the displacement field reduce to:
\begin{subequations}\label{SolBVP2C1meq0}
\begin{equation}\label{SolBVP2C1meq01}
u_r(r,z,t)=\Bigg[\sum_{s=1}^2\bar{A}_s\alpha_sI_{1}(\alpha_sr)\Bigg]\sin\left({k\pi\over L}z\right)\sin(\omega t), 
\end{equation}
\begin{equation}\label{SolBVP2C1meq02}
u_z(r,z,t)=\left({k\pi\over L}\right)\Bigg[\sum_{s=1}^2\bar{A}_s\gamma_sI_0(\alpha_sr)\Bigg]\cos\left({k\pi\over L}z\right)\sin(\omega t),   
\end{equation}
\end{subequations}
where constants $\alpha_s$ and $\gamma_s$ are given by Eqs.~(\ref{alphasEG}) and (\ref{solpsiPP2EG}), respectively, and constants $\left\{\bar{A}_1,\bar{A}_2\right\}$ given by Eq.~(\ref{BVP2C1meq0}).  



\subsection{\textbf{Case 2:} $\displaystyle \left({k\pi\over L}\right)^2<{\rho\omega^2\over(\lambda+2\mu)}<{\rho\omega^2\over\mu}$}

Employing (\ref{GenSolB}) and following the logic at the beginning of Sec.~\ref{BVP1C2}, the displacement components take the form:
\begin{subequations}\label{SolBC2}
\begin{equation}\label{SolBC21}
u_r=\Bigg\{\sum_{s=1}^2\bar{A}_s\bigg[{m\over r}J_m(\alpha_sr)-\alpha_sJ_{m+1}(\alpha_sr)\bigg]+\bar{A}_3{m\over r}J_m(\alpha_2r)\Bigg\}\cos(m\theta)\sin\left({k\pi\over L}z\right)\sin(\omega t), 
\end{equation}
\vspace*{-0.1cm}
\begin{equation}\label{SolBC22}
u_\theta=-\Bigg\{{m\over r}\left(\sum_{s=1}^2\bar{A}_sJ_m(\alpha_sr)\right)+\bar{A}_3\bigg[{m\over r}J_m(\alpha_2r)-\alpha_2J_{m+1}(\alpha_2r)\bigg]\Bigg\}\sin(m\theta)\sin\left({k\pi\over L}z\right)\sin(\omega t),  
\end{equation}
\vspace*{-0.1cm}
\begin{equation}\label{SolBC23}
u_z=\left({k\pi\over L}\right)\Bigg\{\sum_{s=1}^2\bar{A}_s\gamma_sJ_m(\alpha_sr)\Bigg\}\cos(m\theta)\cos\left({k\pi\over L}z\right)\sin(\omega t),   
\end{equation}
\end{subequations}
where constants $\alpha_s$ and $\gamma_s$ are given by Eqs.~(\ref{alphasEG5}) and (\ref{solpsiPP2EG}), respectively. 

The values of the constants $\left\{\bar{A}_1,\bar{A}_2,\bar{A}_3\right\}$ in Eqs.~(\ref{SolBC21})-(\ref{SolBC23}) are again obtained by formal application of boundary conditions (\ref{BCsCRVD2}), which in the present case leads to the $3\times3$ linear system:
\begin{eqnarray}\label{BC2matEQC2}
\left[\begin{array}{ccc}
            F_m+V_{m+1} & G_m+W_{m+1} & (m-1)Q_m-mW_{m+1} \\
            (m-1)P_m-mV_{m+1} & ~(m-1)Q_m-mW_{m+1}~ & H_m+W_{m+1} \\
           2\big(P_m-V_{m+1}\big) & (1+\gamma_2)\big(Q_m-W_{m+1}\big) & Q_m
            \end{array} \right] \left[\begin{array}{c}
            \bar{A}_1 \\
            \bar{A}_2 \\
            \bar{A}_3
           \end{array} \right] =  \left[\begin{array}{r}
           \mathbb{A}~ \\
           -\mathbb{B}~ \\
           \mathbb{C}~    
           \end{array} \right], 
\end{eqnarray}
where the entries in the above $3\times3$ coefficient matrix are given by Eqs.~(\ref{BCmatEQP1C2a})-(\ref{BCmatEQP1C24}), and $\{\mathbb{A},\mathbb{B},\mathbb{C}\}$ are given by (\ref{BCmatEQC15}). The general solution to system (\ref{BC2matEQC2}) is given by Eqs.~(\ref{solnBVP2C1m}), where the coefficients $\left\{\mathsf{C}^{\scriptscriptstyle (i)}_\mathbb{A},\mathsf{C}^{\scriptscriptstyle (i)}_\mathbb{B},\mathsf{C}^{\scriptscriptstyle (i)}_\mathbb{C}:~i=1,2,3\right\}$ are given by Eqs.~(\ref{solnP2C2m2})-(\ref{solnP2C2m10}), and $\mathbb{D}$ is given by Eq.~(\ref{sonP2C2m11}). 

\subsubsection{Special Case: $\displaystyle m=0$}\label{BVP2axysolnC2}

When $m=0$, $u_\theta=0$, $\sigma_{r\theta}=0$, and boundary condition (\ref{strssrt2}) is identically satisfied. Application of boundary conditions (\ref{strssrr2}) and (\ref{strssrz2}) yields the $2\times2$ linear system: 
\begin{eqnarray}\label{BCmatEQC2SCm0}
\left[\begin{array}{cc}
            F_0+V_1 & G_0+W_1 \\
             -2V_1 & ~-\big(1+\gamma_2\big)W_1~
           \end{array} \right] \left[\begin{array}{c}
             \bar{A}_1 \\
             \bar{A}_2
           \end{array} \right] =  \left[\begin{array}{c}
              \mathbb{A} \\
              \mathbb{C}
           \end{array} \right].
\end{eqnarray}
The general solution to system (\ref{BCmatEQC2SCm0}) is:
\begin{subequations}\label{BVP2C2meq0}
\begin{equation}\label{BVPC2meq01}
\bar{A}_1={\big(1+\gamma_2\big)W_1\mathbb{A}+\big(G_0+W_1\big)\mathbb{C}\over\big(1+\gamma_2\big)W_1\big(F_0+V_1\big)-2V_1\big(G_0+W_1\big)},
\end{equation}
\begin{equation}\label{BVPC2meq02}
\bar{A}_2=-{2V_1\mathbb{A}+\big(F_0+V_1\big)\mathbb{C}\over\big(1+\gamma_2\big)W_1\big(F_0+V_1\big)-2V_1\big(G_0+W_1\big)},
\end{equation}
\end{subequations}
where $\left\{F_0,G_0\right\}$ and $\left\{V_1,W_1\right\}$ are the evaluated zero- and first-order Bessel functions (respectively) obtained from substituting $m=0$ in Eqs.~(\ref{BCmatEQP1C2a}), (\ref{BCmatEQP1C2b}), and (\ref{BCmatEQP1C24}). 

Thus, in this special case, the non-zero components of the displacement field reduce to:
\begin{subequations}\label{SolBVP2C2meq0}
\begin{equation}\label{SolBVP2C2meq01}
u_r(r,z,t)=-\Bigg[\sum_{s=1}^2\bar{A}_s\alpha_sJ_{1}(\alpha_sr)\Bigg]\sin\left({k\pi\over L}z\right)\sin(\omega t), 
\end{equation}
\begin{equation}\label{SolBVP2C2meq02}
u_z(r,z,t)=\left({k\pi\over L}\right)\Bigg[\sum_{s=1}^2\bar{A}_s\gamma_sJ_0(\alpha_sr)\Bigg]\cos\left({k\pi\over L}z\right)\sin(\omega t),   
\end{equation}
\end{subequations}
where constants $\alpha_s$ and $\gamma_s$ are given by Eqs.~(\ref{alphasEG5}) and (\ref{solpsiPP2EG}), respectively, and constants $\left\{\bar{A}_1,\bar{A}_2\right\}$ given by Eq.~(\ref{BVP2C2meq0}).  



\subsection{\textbf{Case 3:} $\displaystyle {\rho\omega^2\over(\lambda+2\mu)}<\left({k\pi\over L}\right)^2<{\rho\omega^2\over\mu}$}

Employing (\ref{GenSolB}) and following the logic at the beginning of Sec.~\ref{BVP1C3}, the displacement components take the form:
\begin{subequations}\label{GenSolBC3}
\begin{eqnarray}\label{SolBC31}
u_r&=&\Bigg\{\bar{A}_1\bigg[{m\over r}I_m(\alpha_1r)+\alpha_1I_{m+1}(\alpha_1r)\bigg]+\bar{A}_2\bigg[{m\over r}J_m(\alpha_2r)-\alpha_2J_{m+1}(\alpha_2r)\bigg] \nonumber \\ 
&& \quad +~\bar{A}_3{m\over r}J_m(\alpha_2r)\Bigg\}\cos(m\theta)\sin\left({k\pi\over L}z\right)\sin(\omega t),   
\end{eqnarray}
\vspace*{-0.1cm}
\begin{eqnarray}\label{SolBC32}
u_\theta&=&-\Bigg\{{m\over r}\Big(\bar{A}_1I_m(\alpha_1r)+\bar{A}_2J_m(\alpha_2r)\Big)+\bar{A}_3\bigg[{m\over r}J_m(\alpha_2r)-\alpha_2J_{m+1}(\alpha_2r)\bigg]\Bigg\} \nonumber \\
&& \quad \quad \quad \quad \times \sin(m\theta)\sin\left({k\pi\over L}z\right)\sin(\omega t), 
\end{eqnarray}
\vspace*{-0.1cm}
\begin{equation}\label{SolBC33}
u_z=\left({k\pi\over L}\right)\Bigg\{\bar{A}_1\gamma_1I_m(\alpha_1r)+\bar{A}_2\gamma_2J_m(\alpha_2r)\Bigg\}\cos(m\theta)\cos\left({k\pi\over L}z\right)\sin(\omega t),   
\end{equation}
\end{subequations}
where constants $\alpha_s$ and $\gamma_s$ are given by Eqs.~(\ref{alphasEG3}) and (\ref{solpsiPP2EG}), respectively.

The values of the constants $\left\{\bar{A}_1,\bar{A}_2,\bar{A}_3\right\}$ in Eqs.~(\ref{SolBC31})-(\ref{SolBC33}) are again obtained by formal application of boundary conditions (\ref{BCsCRVD2}), which in the present case leads to the $3\times3$ linear system:
\begin{eqnarray}\label{BC2matEQC3}
\left[\begin{array}{ccc}
            f_m-v_{m+1} & G_m+W_{m+1} & (m-1)Q_m-mW_{m+1} \\
            (m-1)p_m+mv_{m+1} & ~(m-1)Q_m-mW_{m+1}~ & H_m+W_{m+1} \\
           2\big(p_m+v_{m+1}\big) & (1+\gamma_2)\big(Q_m-W_{m+1}\big) & Q_m
            \end{array} \right] \left[\begin{array}{c}
            \bar{A}_1 \\
            \bar{A}_2 \\
            \bar{A}_3
           \end{array} \right] =  \left[\begin{array}{r}
           \mathbb{A}~ \\
           -\mathbb{B}~ \\
           \mathbb{C}~    
           \end{array} \right], 
\end{eqnarray}
where the entries in the above $3\times3$ coefficient matrix are given by Eqs.~(\ref{BCmatEQC12a})-(\ref{BCmatEQC14}) and Eqs.~(\ref{BCmatEQP1C2a})-(\ref{BCmatEQP1C24}), and $\{\mathbb{A},\mathbb{B},\mathbb{C}\}$ are as before given by (\ref{BCmatEQC15}). The general solution to system (\ref{BC2matEQC3}) is given by Eqs.~(\ref{solnBVP2C1m}), where the coefficients $\left\{\mathsf{C}^{\scriptscriptstyle (i)}_\mathbb{A},\mathsf{C}^{\scriptscriptstyle (i)}_\mathbb{B},\mathsf{C}^{\scriptscriptstyle (i)}_\mathbb{C}:~i=1,2,3\right\}$ are given by Eqs.~(\ref{solnP2C3m2})-(\ref{solnP2C3m10}), and $\mathbb{D}$ is given by Eq.~(\ref{sonP2C3m11}). 

\subsubsection{Special Case: $\displaystyle m=0$}\label{BVP2axysolnC3}

When $m=0$, $u_\theta=0$, $\sigma_{r\theta}=0$, and boundary condition (\ref{strssrt2}) is identically satisfied. Application of boundary conditions (\ref{strssrr2}) and (\ref{strssrz2}) yields the $2\times2$ linear system: 
\begin{eqnarray}\label{BCmatEQC3SCm0}
\left[\begin{array}{cc}
            f_0-v_1 & G_0+W_1 \\
             2v_1 & ~-\big(1+\gamma_2\big)W_1~
           \end{array} \right] \left[\begin{array}{c}
             \bar{A}_1 \\
             \bar{A}_2
           \end{array} \right] =  \left[\begin{array}{c}
              \mathbb{A} \\
              \mathbb{C}
           \end{array} \right].
\end{eqnarray}
The general solution to system (\ref{BCmatEQC3SCm0}) is:
\begin{subequations}\label{BVP2C3meq0}
\begin{equation}\label{BVPC3meq01}
\bar{A}_1={\big(1+\gamma_2\big)W_1\mathbb{A}+\big(G_0+W_1\big)\mathbb{C}\over\big(1+\gamma_2\big)W_1\big(f_0-v_1\big)+2v_1\big(G_0+W_1\big)},
\end{equation}
\begin{equation}\label{BVPC3meq02}
\bar{A}_2={2v_1\mathbb{A}-\big(f_0-v_1\big)\mathbb{C}\over\big(1+\gamma_2\big)W_1\big(f_0-v_1\big)+2v_1\big(G_0+W_1\big)},
\end{equation}
\end{subequations}
where $\left\{f_0,G_0\right\}$ and $\left\{v_1,W_1\right\}$ are the evaluated zero- and first-order Bessel functions (respectively) obtained from substituting $m=0$ in Eqs.~(\ref{BCmatEQC12a}), (\ref{BCmatEQP1C2b}), (\ref{BCmatEQC14}), and (\ref{BCmatEQP1C24}). 

Thus, in this special case, the non-zero components of the displacement field reduce to:
\begin{subequations}\label{SolBVP2C3meq0}
\begin{equation}\label{SolBVP2C3meq01}
u_r(r,z,t)=\Bigg[\bar{A}_1\alpha_1I_{1}(\alpha_1r)-\bar{A}_2\alpha_2J_{1}(\alpha_2r)\Bigg]\sin\left({k\pi\over L}z\right)\sin(\omega t), 
\end{equation}
\begin{equation}\label{SolBVP2C3meq02}
u_z(r,z,t)=\left({k\pi\over L}\right)\Bigg[\bar{A}_1\gamma_1I_0(\alpha_1r)+\bar{A}_2\gamma_2J_0(\alpha_2r)\Bigg]\cos\left({k\pi\over L}z\right)\sin(\omega t), 
\end{equation}
\end{subequations}
where constants $\alpha_s$ and $\gamma_s$ are given by Eqs.~(\ref{alphasEG3}) and (\ref{solpsiPP2EG}), respectively, and constants $\left\{\bar{A}_1,\bar{A}_2\right\}$ given by Eq.~(\ref{BVP2C3meq0}).  



\subsection{Consistency with the ENBKS Field Equations}\label{consisSOLNs}

As an analytical check, we have verified that, in the special $m=0$ case, the general stress and displacement fields obtained from applying our method of solution to BVP 2 are consistent with the general axisymmetric field equations that would be obtained from applying the formalism of Ebenezer et al. \cite{Ebenezer08} (henceforth referred to as the ENBKS method) to BVP 2. Demonstration of this consistency is somewhat intricate; the details are therefore consigned to Appendix \ref{appendx2}. Equivalency of our $m=0$ solution with the axisymmetric solution that would be obtained from the ENBKS method then directly follows from application of the boundary conditions.  

\section{Numerical Example}

\begin{table}[h]
\centering
\begin{tabular}{ll}
    \hline \hline 
     Parameter & Numerical Value \\ \hline
     Length ($L$) & $0.15$ m \\ 
     Radius ($R$) & $0.05$ m  \\ 
     Mass density ($\rho$) & $8000~\text{kg/m}^3$  \\ 
     Young modulus ($E$)  & $190$ GPa \\ 
     Poisson ratio ($\nu$)  & $0.30$  \\ 
     First Lam\'{e} constant ($\lambda$)  & $109.62$ GPa \\ 
     Second Lam\'{e} constant ($\mu$) \hspace*{2cm} & $73.08$ GPa \\ \hline \hline
\end{tabular}
\caption{Geometric and material parameter values used in the numerical example.}
\label{GeoMatParams}
\end{table}

\begin{table}[h]
\begin{tabular}{ccccccccc}
    \hline \hline 
    Mode number &~~~~~& \multicolumn{7}{c}{Circumferential wave number} \\ 
    \cline{3-9} 
    && $m=0$ &~~~~~& $m=1$ &~~~~~& $m=2$ &~~~~~& $m=3$ \\ \hline
    1 && 15.766 &~~~~~&   6.118 &~~~~~& 22.382 &~~~~~& 34.557 \\
    2 && 27.063 &~~~~~& 16.217 &~~~~~& 25.540 &~~~~~& 36.163 \\
    3 && 33.537 &~~~~~& 17.713 &~~~~~& 29.383 &~~~~~& 40.417 \\
    4 && 34.268 &~~~~~& 23.700 &~~~~~& 32.057 &~~~~~& 40.423 \\
    5 && 36.863 &~~~~~& 26.495 &~~~~~& 33.123 &~~~~~& 42.922 \\
    6 && 39.808 &~~~~~& 29.733 &~~~~~& 39.233 &~~~~~& 46.640 \\
    7 && 40.560 &~~~~~& 31.124 &~~~~~& 40.188 &~~~~~& 47.822 \\
    8 && 45.241 &~~~~~& 36.568 &~~~~~& 44.487 &~~~~~& 53.510 \\
    9 && 48.525 &~~~~~& 38.137 &~~~~~& 45.590 &~~~~~& 54.054 \\ \hline \hline     
\end{tabular}
\caption{Natural frequencies of a simply-supported isotropic solid elastic circular cylinder having the geometrical and material properties given in Table \ref{GeoMatParams}. All values are in units of kHz. The above data was computed using the free-vibration frequency data given in Table 7 of Ref.~\cite{Ye14}.}
\label{TabNatFreqs}
\end{table}

\begin{figure}[h]
\vspace*{-0.9cm} 
\centering 
\scalebox{0.6}{\includegraphics*{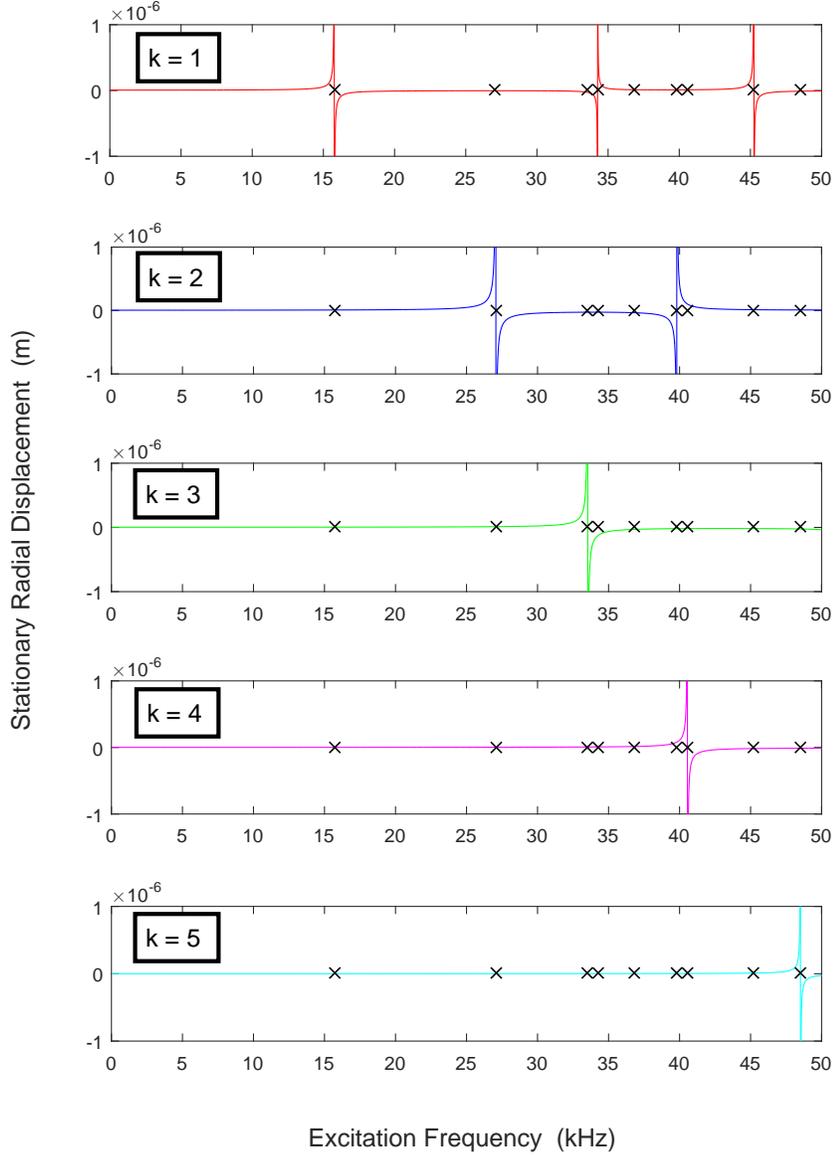}}
\vspace*{-0.85cm}
\caption{\label{radfreqresmeq0} Stationary radial displacement at point $(r,\theta,z)=(R/2,0,L/7)$ for various values of the longitudinal wave number $k$ and circumferential wave number $m=0$. For reference, the natural frequencies listed in Table \ref{TabNatFreqs} are marked by `$\mathsf{X}$'s on the frequency axis of each subplot.}
\end{figure}

\begin{figure}[h]
\vspace*{-0.9cm} 
\centering 
\scalebox{0.6}{\includegraphics*{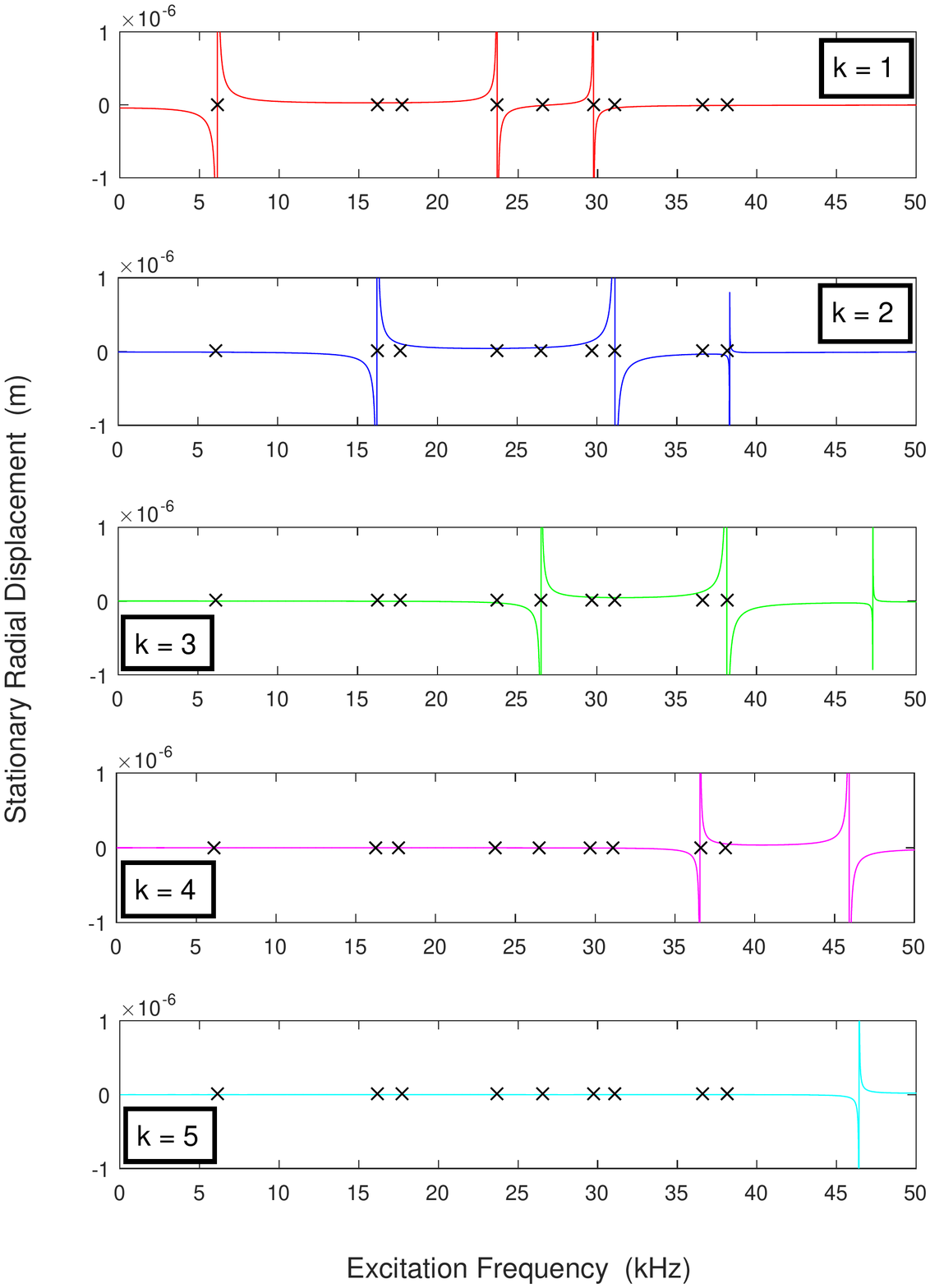}}
\vspace*{-0.85cm}
\caption{\label{radfreqresmeq1} Same as Figure \ref{radfreqresmeq0} except the circumferential wave number $m=1$.}
\end{figure}

\begin{figure}[h]
\vspace*{-0.9cm} 
\centering 
\scalebox{0.6}{\includegraphics*{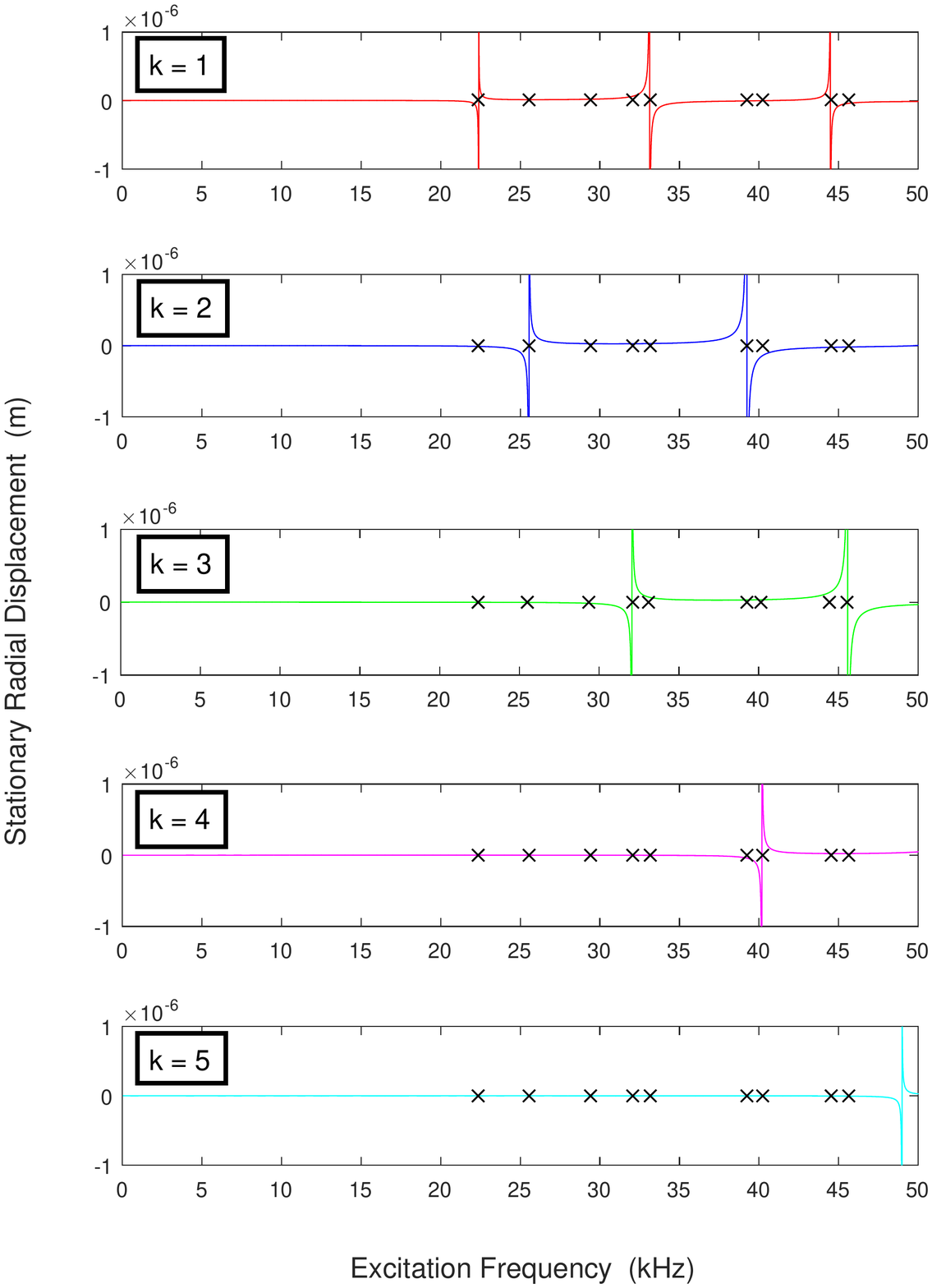}}
\vspace*{-0.85cm}
\caption{\label{radfreqresmeq2} Same as Figure \ref{radfreqresmeq0} except the circumferential wave number $m=2$.}
\end{figure}

\begin{figure}[h]
\vspace*{-0.9cm} 
\centering 
\scalebox{0.475}{\includegraphics*{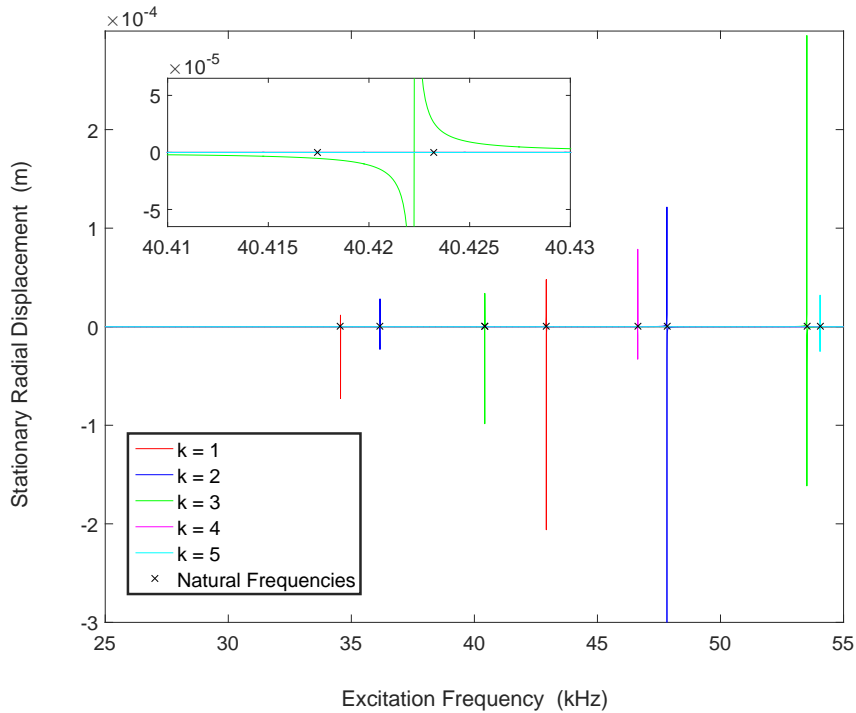}}
\vspace*{-0.7cm}
\caption{\label{radfreqresmeq3} Responses at point $(r,\theta,z)=(R/2,0,L/7)$ to various standing-wave excitations with longitudinal wave numbers $k$ as indicated and circumferential wave number $m=3$. As discussed in the text, the inset shows a closer view of the neighborhood around the third resonance. A near-degeneracy in the natural frequency spectrum occurs inside the interval $(40,41)$ kHz.}
\end{figure}

\begin{figure}[h]
\vspace*{-0.9cm} 
\centering 
\scalebox{0.6}{\includegraphics*{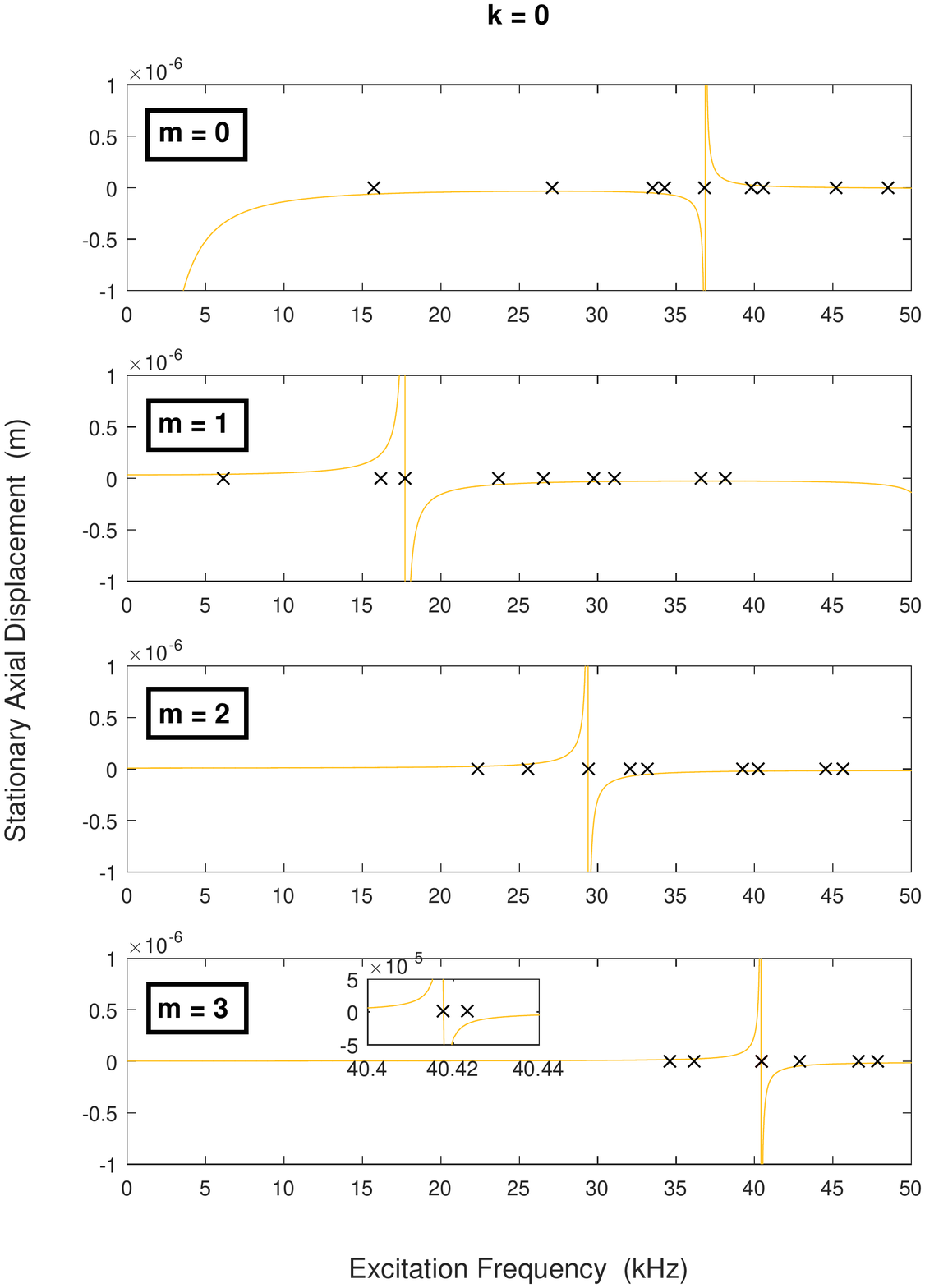}}
\vspace*{-0.9cm}
\caption{\label{freqreskeq0} Stationary axial displacement at point $(r,\theta,z)=(R/2,0,L/7)$ for various values of the circumferential wave number $m$ and longitudinal wave number $k=0$. For reference, the natural frequencies listed in Table \ref{TabNatFreqs} are marked by `$\mathsf{X}$'s on the frequency axis of each subplot.}
\end{figure} 

As an example, we examine the steady-state forced-vibration response of a solid steel cylinder subjected to harmonic standing-wave excitations of type (\ref{BCsCRVD2}) on its curved surface. The geometric and material properties of the cylinder are specified in Table \ref{GeoMatParams}. While there are several possible types of analyses that could be considered in studying the steady-state forced response, the most common is to study the behavior of the stationary displacement as a function of the excitation frequency. We shall here restrict attention to this type of analysis. Recall that the stationary displacement in the context of steady-state (i.e., time-harmonic) vibrations is merely the time-independent part of the displacement field. While this is a 3D function of the spatial coordinates and thus varies from point to point in the cylinder, its qualitative global behavior as a function of the excitation frequency should generally be independent of location in the cylinder. This will certainly be true under general variations in the circumferential and longitudinal coordinates since the corresponding parts of the displacement field are independent of excitation frequency. Variations in the radial coordinate are effectively scale transformations of the otherwise frequency-dependent Bessel functions. In `frequency space', the effects of such transformations can be significant locally but not so  globally for the simple reason that all physically relevant features in frequency space are due to the boundary conditions and are thereby mainly determined by the frequency-dependent amplitudes $\left\{\bar{A}_1,\bar{A}_2,\bar{A}_3\right\}$. As is usual for this type of analysis, we shall thus study the stationary part of the displacement field, at a few suitably-chosen representative points in the cylinder, as a function of the excitation frequency\footnote{Note that, aside from nodal points, we are free to choose any point in the cylinder as a representative point.}. To do so, we shall numerically evaluate the analytical solution to BVP 2 obtained in Sec.~\ref{solntoBVP2}. 

Using parameter values as specified in Table \ref{GeoMatParams}, the components of the displacement field were computed at excitation frequencies that are integer multiples of 10 Hz with lower and upper bounds of 10 Hz and 100 kHz, respectively. In all calculations, the excitation amplitudes were set as follows: $\mathcal{A}=\mathcal{B}=\mathcal{C}=10^5$ Pa. Given that the cylinder is being forced to vibrate, we expect to observe unusually large displacements (i.e., resonances) when the excitation frequency is close to one of the natural frequencies of the \emph{simply-supported} cylinder. The first nine of these frequencies, computed using free-vibration frequency data obtained from Ref.~\cite{Ye14}, are given in Table \ref{TabNatFreqs}. The (stationary) radial displacement at the representative interior point $(r=R/2,\theta=0,z=L/7)$ is shown in Figs.~\ref{radfreqresmeq0}-\ref{radfreqresmeq2} for various values of the longitudinal wave number $k$ and circumferential wave numbers $m=0$, 1, and 2, respectively. For visual reference, the natural frequencies listed in Table \ref{TabNatFreqs} are marked by `$\mathsf{X}$'s on the frequency axis of each subplot.

As seen in Figs.~\ref{radfreqresmeq0}-\ref{radfreqresmeq2} and as well observed in other more general forced-vibration analyses of solid elastic cylinders \cite{Ebenezer05,Ebenezer08}, the response at a particular frequency depends on the spatial distribution of the excitation, which is here determined by the specific values of the wave numbers $m$ and $k$. In each case, we observe unmistakable resonances close to (a subset of) the natural frequencies of the simply-supported cylinder. Note that each \emph{individual} excitation (obtained by specifying a single pair of $(m,k)$ values) generates a unique \emph{series} of resonances, as opposed to producing just a single resonance. In other words, a single harmonic excitation excites a set of resonant modes instead of exciting only one resonant mode. Unfortunately, there does not appear to be any simple rule for predicting which resonant modes will be excited by a particular excitation. 

In Figs.~\ref{radfreqresmeq0}-\ref{radfreqresmeq2}, we used common vertical and horizontal scales in all subplots in order to make it easier to compare the different cases. We should however mention that the amplitudes of the resonances are not all equal, and this is evident when one views the response outside the displacement range shown in the figures. This feature is illustrated in Fig.~\ref{radfreqresmeq3}, which shows the responses to the five lowest $k$-harmonic standing-wave excitations [at the representative point $(r=R/2,\theta=0,z=L/7)$] when $m=3$. (Higher $k$-harmonics excite resonant modes at higher excitation frequencies than those considered here.) Note that the displacements generated by these five different standing-wave excitations $\{(m,k): m=3, k=1,\ldots,5\}$ are overlaid on the same plot with the understanding that they actually correspond to different excitation cases. So, it should be understood that, for instance, the red curve is the response to a standing-wave excitation with longitudinal wave number $k=1$, whereas the green curve is the response to a standing-wave excitation with longitudinal wave number $k=3$. The natural frequencies listed in the last column of Table \ref{TabNatFreqs} are again marked by `$\mathsf{X}$'s on the frequency axis. No resonances occur in the frequency interval $(0,25)$ kHz. For this reason, we only display the result on the frequency interval $[25,55]$ kHz. It is clear from this plot, which is representative, that the amplitudes of the resonances generally differ. Note that differences in amplitude not only occur between the different excitation cases; the amplitudes of the resonances generated by each individual excitation also vary. While it may be obvious to some readers, it is worth emphasizing that the amplitudes, which inherently depend on the frequency resolution\footnote{In the main plot of Fig.~\ref{radfreqresmeq3} (i.e., not the inset), the components of the displacement field were computed at excitation frequencies that are integer multiples of 1 Hz with lower and upper bounds of 1 Hz and 100 kHz, respectively. When a different frequency increment is used (e.g., 0.5 Hz instead of 1 Hz), the amplitudes change.} and on where in the cylinder the response is determined, should not be interpreted as resonance intensities. In the present context of a lossless (i.e., undamped) cylinder, the amplitudes are insignificant since, in theory, the amplitude of any resonance asymptotically approaches infinity as the excitation frequency approaches the associated natural frequency. Resonance intensities are actually related (non-trivially) to the shapes and widths of the resonances, and thus meaningful comparisons of resonance intensities can be made by careful examination of the resonance widths. Accurately calculating resonance intensities is a specialized subject in its own right and one that is more appropriately discussed when there are material damping effects. We shall thus not consider this topic any further here. With respect to studying the elastodynamic response as a function of excitation frequency, our intent here is only to illustrate and discuss the results of numerically evaluating our analytical solution. 

Near-degeneracies in the natural frequency spectrum are not uncommon. (Note that we are here specifically referring to the situation in which two or more consecutive natural frequencies have nearly equal values.) This occurs, for example, in the low end of the frequency spectrum when $m=3$. As can be seen from the last column of Table \ref{TabNatFreqs}, the third and fourth natural frequencies differ by less than 10 Hz, which is a small difference on a kHz scale. Such situations are interesting in the context of forced vibrations since it is not generally obvious how many individual resonances will occur or can be resolved when the excitation frequency is in the vicinity of several closely-spaced natural frequencies. In theory, one generally expects to observe one resonance per natural frequency, but for a variety of reasons, this does not always occur in practice. If fewer resonances occur than expected, then it is informative to determine their relative intensities. Again, it does not suit our purpose  to delve into this topic here. We merely want to point out that no further individually-resolved resonances are revealed when zooming in on the third resonance in Fig.~\ref{radfreqresmeq3}. To be certain, we re-computed the components of the displacement field at excitation frequencies in the interval $[40,41]$ kHz in increments of 0.1 Hz. (A frequency increment of 1 Hz was employed in obtaining the main plot of Fig.~\ref{radfreqresmeq3}.) A zoomed view of the result is shown in the inset of Fig.~\ref{radfreqresmeq3} from which it is clear that there is no resonance associated with the third natural frequency. 

In Figs.~\ref{radfreqresmeq0}-\ref{radfreqresmeq3}, the absence of resonances in the neighborhoods of certain natural frequencies, in particular, the fifth natural frequency when $m=0$ and the third natural frequency when $m=\{1,2,3\}$ is noteworthy and interesting. As it turns out, resonant modes at these frequencies are excited by harmonic boundary stresses of type (\ref{BCsCRVD2}) with longitudinal wave number $k=0$, as shown in Fig.~\ref{freqreskeq0}. The analytical solutions obtained in Secs.~\ref{solntoBVP1} and \ref{solntoBVP2} are valid only when $k$ is a positive integer. Analytical solutions to BVP 1 and BVP 2 can however also be obtained when $k=0$. The solution to BVP 2 is given in Appendix \ref{appendx}. The results of Figs.~\ref{radfreqresmeq0}-\ref{freqreskeq0} therefore imply that boundary stresses of the harmonic standing-wave type excite resonant modes at all natural frequencies. 

Similar results are obtained for all components of the displacement field evaluated at any suitably-chosen representative point. Numerical experiments bear out that, regardless of displacement component or location in the cylinder, the frequency response is qualitatively always the same; each standing-wave excitation generates a series of resonances that are in correspondence with a unique subset of the natural frequency spectrum of the simply-supported cylinder. Quantitative differences in the detailed features of the resonances (e.g., their shapes and widths) are observed when results for the different displacement components are overlaid or when the representative point is varied, but these differences are not usually of interest in steady-state vibration analyses of undamped solid elastic cylinders \cite{Ebenezer05,Ebenezer08}. 

\section{Conclusion}

In this paper, we considered the problem of determining the steady-state forced-vibration response of a simply-supported isotropic solid elastic circular cylinder subjected to 2D harmonic standing-wave excitations on its curved surface. Exploiting previously-obtained particular solutions to the Navier-Lam\'{e} equation \cite{usB1} and exact matrix algebra, we constructed an exact closed-form 3D elastodynamic solution. In constructing the solution, we made use of standard Bessel function identities in order to cast the radial part of the solution, which involves Bessel functions of the first kind, in a symmetric form that is free of derivatives thereby making the obtained analytical solution apt for numerical computation. 

Two complete analytical solutions were in fact constructed corresponding to two different but closely-related families of harmonic standing-wave excitations. The second of these analytical solutions (i.e., the solution to BVP 2) was evaluated numerically in order to study the frequency response in some example cases. In each case, the solution generates a series of resonances in correspondence with (a subset of) the natural frequencies of the \emph{simply-supported} cylinder. It is worth emphasizing that each \emph{individual} standing-wave excitation generates a unique \emph{series} of resonances, as opposed to producing just a single resonance. In general, the response at a particular frequency depends on the spatial distribution of the excitation, which is, in the present context, determined by the specific values of the wave numbers $m$ and $k$. This is consistent with what is observed in other forced-response studies of solid elastic cylinders \cite{Ebenezer05,Ebenezer08}. Finally, we note that while harmonic standing-wave excitations of type (\ref{BCsCRVD1}) and (\ref{BCsCRVD2}) excite resonant modes at all natural frequencies, there does not appear to be any simple rule for predicting which resonant modes will be excited by a particular excitation. 

The problem considered in this paper is of general interest both as an exactly-solvable 3D elastodynamics problem and as a benchmark forced-vibration problem involving a solid elastic cylinder. The obtained analytical solution is not only useful for revealing the main physical features of the considered problem, but can also serve as a benchmark solution for assessment and/or validation purposes. The method of solution employed in this paper demonstrates a general approach that can be applied to solve other elastodynamic forced-response problems involving isotropic, open or closed, solid or hollow, elastic cylinders with simply-supported or other boundary conditions. 

\begin{acknowledgments}
The authors acknowledge financial support from the Natural Sciences and Engineering Research Council (NSERC) of Canada and the Ontario Research Foundation (ORF). 
\end{acknowledgments}

\appendix

\section{Exact Values of the Solution Coefficients $\left\{\mathsf{C}^{\scriptscriptstyle (i)}_\mathbb{A},\mathsf{C}^{\scriptscriptstyle (i)}_\mathbb{B},\mathsf{C}^{\scriptscriptstyle (i)}_\mathbb{C}:~i=1,2,3\right\}$}\label{solncoeffs}

\subsection{Case 1}\label{solncoeffsC1}

\begin{subequations}\label{solncoeffsC1are}
\begin{equation}\label{solnP2C1m2}
\mathsf{C}^{\scriptscriptstyle (1)}_\mathbb{A}=\big((m-1)q_m+mw_{m+1}\big)q_m-(1+\gamma_2)\big(h_m-w_{m+1}\big)\big(q_m+w_{m+1}\big),
\end{equation}
\begin{equation}\label{solnP2C1m3}
\mathsf{C}^{\scriptscriptstyle (1)}_\mathbb{B}=(1+\gamma_2)\big((m-1)q_m+mw_{m+1}\big)\big(q_m+w_{m+1}\big)-\big(g_m-w_{m+1}\big)q_m,
\end{equation}
\begin{equation}\label{solnP2C1m4}
\mathsf{C}^{\scriptscriptstyle (1)}_\mathbb{C}=\big(g_m-w_{m+1}\big)\big(h_m-w_{m+1}\big)-\big((m-1)q_m+mw_{m+1}\big)\big((m-1)q_m+mw_{m+1}\big),
\end{equation}
\begin{equation}\label{solnP2C1m5}
\mathsf{C}^{\scriptscriptstyle (2)}_\mathbb{A}=\big((m-1)p_m+mv_{m+1}\big)q_m-2\big(h_m-w_{m+1}\big)\big(p_m+v_{m+1}\big),
\end{equation}
\begin{equation}\label{solnP2C1m6}
\mathsf{C}^{\scriptscriptstyle (2)}_\mathbb{B}=2\big((m-1)q_m+mw_{m+1}\big)\big(p_m+v_{m+1}\big)-\big(f_m-v_{m+1}\big)q_m,
\end{equation}
\begin{equation}\label{solnP2C1m7}
\mathsf{C}^{\scriptscriptstyle (2)}_\mathbb{C}=\big(f_m-v_{m+1}\big)\big(h_m-w_{m+1}\big)-\big((m-1)q_m+mw_{m+1}\big)\big((m-1)p_m+mv_{m+1}\big),
\end{equation}
\begin{equation}\label{solnP2C1m8}
\mathsf{C}^{\scriptscriptstyle (3)}_\mathbb{A}=(1+\gamma_2)\big((m-1)p_m+mv_{m+1}\big)\big(q_m+w_{m+1}\big)-2\big((m-1)q_m+mw_{m+1}\big)\big(p_m+v_{m+1}\big),
\end{equation}
\begin{equation}\label{solnP2C1m9}
\mathsf{C}^{\scriptscriptstyle (3)}_\mathbb{B}=2\big(g_m-w_{m+1}\big)\big(p_m+v_{m+1}\big)-(1+\gamma_2)\big(f_m-v_{m+1}\big)\big(q_m+w_{m+1}\big),
\end{equation}
\begin{equation}\label{solnP2C1m10}
\mathsf{C}^{\scriptscriptstyle (3)}_\mathbb{C}=\big(f_m-v_{m+1}\big)\big((m-1)q_m+mw_{m+1}\big)-\big(g_m-w_{m+1}\big)\big((m-1)p_m+mv_{m+1}\big),
\end{equation}
and
\begin{eqnarray}\label{sonP2C1m11}
&&\hspace*{-0.5cm}\mathbb{D}=2\Big[\big((m-1)q_m+mw_{m+1}\big)^2-\big(g_m-w_{m+1}\big)(h_m-w_{m+1}\big)\Big]\big(p_m+v_{m+1}\big) \nonumber \\ 
&&\hspace*{-0.5cm}+~(1+\gamma_2)\Big[\big(f_m-v_{m+1}\big)\big(h_m-w_{m+1}\big)-\big((m-1)q_m+mw_{m+1}\big)\big((m-1)p_m+mv_{m+1}\big)\Big]\big(q_m+w_{m+1}\big) \nonumber \\ 
&&\hspace*{-0.5cm}+~\Big[\big(g_m-w_{m+1}\big)\big((m-1)p_m+mv_{m+1}\big)-\big(f_m-v_{m+1}\big)\big((m-1)q_m+mw_{m+1}\big)\Big]q_m. 
\end{eqnarray}
\end{subequations}

\subsection{Case 2}\label{solncoeffsC2}

\begin{subequations}\label{solncoeffsC2are}
\begin{equation}\label{solnP2C2m2}
\mathsf{C}^{\scriptscriptstyle (1)}_\mathbb{A}=\big((m-1)Q_m-mW_{m+1}\big)Q_m-(1+\gamma_2)\big(H_m+W_{m+1}\big)\big(Q_m-W_{m+1}\big),
\end{equation}
\begin{equation}\label{solnP2C2m3}
\mathsf{C}^{\scriptscriptstyle (1)}_\mathbb{B}=(1+\gamma_2)\big((m-1)Q_m-mW_{m+1}\big)\big(Q_m-W_{m+1}\big)-\big(G_m+W_{m+1}\big)Q_m,
\end{equation}
\begin{equation}\label{solnP2C2m4}
\mathsf{C}^{\scriptscriptstyle (1)}_\mathbb{C}=\big(G_m+W_{m+1}\big)\big(H_m+W_{m+1}\big)-\big((m-1)Q_m-mW_{m+1}\big)\big((m-1)Q_m-mW_{m+1}\big),
\end{equation}
\begin{equation}\label{solnP2C2m5}
\mathsf{C}^{\scriptscriptstyle (2)}_\mathbb{A}=\big((m-1)P_m-mV_{m+1}\big)Q_m-2\big(H_m+W_{m+1}\big)\big(P_m-V_{m+1}\big),
\end{equation}
\begin{equation}\label{solnP2C2m6}
\mathsf{C}^{\scriptscriptstyle (2)}_\mathbb{B}=2\big((m-1)Q_m-mW_{m+1}\big)\big(P_m-V_{m+1}\big)-\big(F_m+V_{m+1}\big)Q_m,
\end{equation}
\begin{equation}\label{solnP2C2m7}
\mathsf{C}^{\scriptscriptstyle (2)}_\mathbb{C}=\big(F_m+V_{m+1}\big)\big(H_m+W_{m+1}\big)-\big((m-1)Q_m-mW_{m+1}\big)\big((m-1)P_m-mV_{m+1}\big),
\end{equation}
\begin{equation}\label{solnP2C2m8}
\mathsf{C}^{\scriptscriptstyle (3)}_\mathbb{A}=(1+\gamma_2)\big((m-1)P_m-mV_{m+1}\big)\big(Q_m-W_{m+1}\big)-2\big((m-1)Q_m-mW_{m+1}\big)\big(P_m-V_{m+1}\big),
\end{equation}
\begin{equation}\label{solnP2C2m9}
\mathsf{C}^{\scriptscriptstyle (3)}_\mathbb{B}=2\big(G_m+W_{m+1}\big)\big(P_m-V_{m+1}\big)-(1+\gamma_2)\big(F_m+V_{m+1}\big)\big(Q_m-W_{m+1}\big),
\end{equation}
\begin{equation}\label{solnP2C2m10}
\mathsf{C}^{\scriptscriptstyle (3)}_\mathbb{C}=\big(F_m+V_{m+1}\big)\big((m-1)Q_m-mW_{m+1}\big)-\big(G_m+W_{m+1}\big)\big((m-1)P_m-mV_{m+1}\big),
\end{equation}
and
\begin{eqnarray}\label{sonP2C2m11}
&&\hspace*{-0.5cm}\mathbb{D}=2\Big[\big((m-1)Q_m-mW_{m+1}\big)^2-\big(G_m+W_{m+1}\big)\big(H_m+W_{m+1}\big)\Big]\big(P_m-V_{m+1}\big) \nonumber \\ 
&&\hspace*{-0.5cm}+~(1+\gamma_2)\Big[\big(F_m+V_{m+1}\big)\big(H_m+W_{m+1}\big)-\big((m-1)Q_m-mW_{m+1}\big)\big((m-1)P_m-mV_{m+1}\big)\Big]\big(Q_m-W_{m+1}\big) \nonumber \\ 
&&\hspace*{-0.5cm}+~\Big[\big(G_m+W_{m+1}\big)\big((m-1)P_m-mV_{m+1}\big)-\big(F_m+V_{m+1}\big)\big((m-1)Q_m-mW_{m+1}\big)\Big]Q_m. 
\end{eqnarray}
\end{subequations} 

\subsection{Case 3}\label{solncoeffsC3}

\begin{subequations}\label{solncoeffsC3are}
\begin{equation}\label{solnP2C3m2}
\mathsf{C}^{\scriptscriptstyle (1)}_\mathbb{A}=\big((m-1)Q_m-mW_{m+1}\big)Q_m-(1+\gamma_2)\big(H_m+W_{m+1}\big)\big(Q_m-W_{m+1}\big),
\end{equation}
\begin{equation}\label{solnP2C3m3}
\mathsf{C}^{\scriptscriptstyle (1)}_\mathbb{B}=(1+\gamma_2)\big((m-1)Q_m-mW_{m+1}\big)\big(Q_m-W_{m+1}\big)-\big(G_m+W_{m+1}\big)Q_m,
\end{equation}
\begin{equation}\label{solnP2C3m4}
\mathsf{C}^{\scriptscriptstyle (1)}_\mathbb{C}=\big(G_m+W_{m+1}\big)\big(H_m+W_{m+1}\big)-\big((m-1)Q_m-mW_{m+1}\big)\big((m-1)Q_m-mW_{m+1}\big),
\end{equation}
\begin{equation}\label{solnP2C3m5}
\mathsf{C}^{\scriptscriptstyle (2)}_\mathbb{A}=\big((m-1)p_m+mv_{m+1}\big)Q_m-2\big(H_m+W_{m+1}\big)\big(p_m+v_{m+1}\big),
\end{equation}
\begin{equation}\label{solnP2C3m6}
\mathsf{C}^{\scriptscriptstyle (2)}_\mathbb{B}=2\big((m-1)Q_m-mW_{m+1}\big)\big(p_m+v_{m+1}\big)-\big(f_m-v_{m+1}\big)Q_m,
\end{equation}
\begin{equation}\label{solnP2C3m7}
\mathsf{C}^{\scriptscriptstyle (2)}_\mathbb{C}=\big(f_m-v_{m+1}\big)\big(H_m+W_{m+1}\big)-\big((m-1)Q_m-mW_{m+1}\big)\big((m-1)p_m+mv_{m+1}\big),
\end{equation}
\begin{equation}\label{solnP2C3m8}
\mathsf{C}^{\scriptscriptstyle (3)}_\mathbb{A}=(1+\gamma_2)\big((m-1)p_m+mv_{m+1}\big)\big(Q_m-W_{m+1}\big)-2\big((m-1)Q_m-mW_{m+1}\big)\big(p_m+v_{m+1}\big),
\end{equation}
\begin{equation}\label{solnP2C3m9}
\mathsf{C}^{\scriptscriptstyle (3)}_\mathbb{B}=2\big(G_m+W_{m+1}\big)\big(p_m+v_{m+1}\big)-(1+\gamma_2)\big(f_m-v_{m+1}\big)\big(Q_m-W_{m+1}\big),
\end{equation}
\begin{equation}\label{solnP2C3m10}
\mathsf{C}^{\scriptscriptstyle (3)}_\mathbb{C}=\big(f_m-v_{m+1}\big)\big((m-1)Q_m-mW_{m+1}\big)-\big(G_m+W_{m+1}\big)\big((m-1)p_m+mv_{m+1}\big),
\end{equation}
and
\begin{eqnarray}\label{sonP2C3m11}
&&\hspace*{-0.5cm}\mathbb{D}=2\Big[\big((m-1)Q_m-mW_{m+1}\big)^2-\big(G_m+W_{m+1}\big)\big(H_m+W_{m+1}\big)\Big]\big(p_m+v_{m+1}\big) \nonumber \\ 
&&\hspace*{-0.5cm}+~(1+\gamma_2)\Big[\big(f_m-v_{m+1}\big)\big(H_m+W_{m+1}\big)-\big((m-1)Q_m-mW_{m+1}\big)\big((m-1)p_m+mv_{m+1}\big)\Big]\big(Q_m-W_{m+1}\big) \nonumber \\ 
&&\hspace*{-0.5cm}+~\Big[\big(G_m+W_{m+1}\big)\big((m-1)p_m+mv_{m+1}\big)-\big(f_m-v_{m+1}\big)\big((m-1)Q_m-mW_{m+1}\big)\Big]Q_m. 
\end{eqnarray}
\end{subequations} 

\section{Solution to BVP 2 in the Special Case $\displaystyle k=0$}\label{appendx}

It can be established (employing results from Ref.~\cite{usB1} or otherwise) that 
\begin{equation}\label{gensolnkeq0}
u_r=0, \quad u_\theta=0, \quad u_z=AJ_m(\alpha r)\cos(m\theta)\sin(\omega t), 
\end{equation}
where $\alpha=\sqrt{\rho\omega^2/\mu}$, is a particular solution to Eq.~(\ref{NLE}) compatible with boundary conditions (\ref{strssrr2})-(\ref{strssrz2}) when $k=0$. Solution (\ref{gensolnkeq0}) therefore furnishes the general form of the displacement field appropriate to the boundary-value problem defined in Sec.~\ref{EGFULL} in the special case $k=0$. We need now only to determine the value of the constant $A$ in (\ref{gensolnkeq0}) that satisfies boundary conditions (\ref{strssrr2})-(\ref{strssrz2}). Substituting the components of solution (\ref{gensolnkeq0}) into Eqs.~(\ref{stssstrnCYL1}), (\ref{stssstrnCYL4}), and (\ref{stssstrnCYL5}) immediately yields the stress components:
\begin{subequations}\label{strsssolnkeq0}
\begin{equation}\label{strsssolnkeq0A}
\sigma_{rr}(r,\theta,z,t)=\sigma_{r\theta}(r,\theta,z,t)=0,
\end{equation}
and
\begin{equation}\label{strsssolnkeq0B}
\sigma_{rz}(r,\theta,t)=\mu A\bigg[{m\over r}J_m(\alpha r)-\alpha J_{m+1}(\alpha r)\bigg]\cos(m\theta)\sin(\omega t).
\end{equation}
\end{subequations}
Boundary conditions (\ref{strssrr2}) and (\ref{strssrt2}) are therefore satisfied identically. Application of boundary condition (\ref{strssrz2}) with $k=0$ yields a condition involving the constant $A$ that can be subsequently solved to give
\begin{equation}\label{solnkeq0A}
A={\mathcal{C}\over\mu\left[{m\over R}J_m(\alpha R)-\alpha J_{m+1}(\alpha R)\right]}.
\end{equation}

\section{Axisymmetric Solution to BVP 2 Using the ENBKS Method}\label{appendx2}

In this appendix, we show how the axisymmetric solution to BVP 2 may be reproduced using the ENBKS method. We shall here only consider Case 1. The solution in the other two cases may be similarly reproduced.  

\subsection{General ENBKS Solution}

Quoting from Ref.~\cite{Ebenezer08}, the following is an exact \emph{axisymmetric} solution to Eq.~(\ref{NLE}), for arbitrary values of $k_{rn}~(n=1,2,3,\ldots,N_r)$ and $k_{zn}~(n=1,2,3,\ldots,N_z)$: 
\begin{subequations}\label{Ebensoln}
\begin{eqnarray}\label{EbensolnT}
\left[\begin{array}{c} u^{(ENBKS)}_z \\ u^{(ENBKS)}_r \end{array} \right] = \left[\begin{array}{c}
u^{(1)}_z \\ u^{(1)}_r \end{array} \right] + \left[\begin{array}{c}
u^{(2)}_z \\ u^{(2)}_r \end{array} \right], 
\end{eqnarray}
where
\begin{eqnarray}\label{EbensolnP1}
\left[\begin{array}{c} u^{(1)}_z \\ u^{(1)}_r \end{array} \right] = \left[\begin{array}{c} P\cos(K_1z)+{\displaystyle\sum_{n=1}^{N_r}}{\displaystyle\sum_{s=1}^2}P_{ns}J_0(k_{rn}r)\cos(k_{zns}z) \\ {\displaystyle\sum_{n=1}^{N_r}}{\displaystyle\sum_{s=1}^2}P_{ns}\psi_{ns}J_1(k_{rn}r)\sin(k_{zns}z) \end{array} \right],  
\end{eqnarray}
\begin{eqnarray}\label{EbensolnP2}
\left[\begin{array}{c} u^{(2)}_z \\ u^{(2)}_r \end{array} \right] = \left[\begin{array}{c} QJ_0(K_2r)+{\displaystyle\sum_{n=1}^{N_z}}{\displaystyle\sum_{s=1}^2}Q_{ns}J_0(k_{rns}r)\cos(k_{zn}z) \\ {\displaystyle\sum_{n=1}^{N_z}}{\displaystyle\sum_{s=1}^2}Q_{ns}\chi_{ns}J_1(k_{rns}r)\sin(k_{zn}z) \end{array} \right],  
\end{eqnarray}
\begin{equation}
K_1=\sqrt{\rho\omega^2\over(\lambda+2\mu)}, \quad K_2=\sqrt{\rho\omega^2\over\mu},
\end{equation}
$P$, $Q$, $\{P_{ns}:n=1,2,3,\ldots,N_r;~s=1,2\}$, $\{Q_{ns}:n=1,2,3,\ldots,N_z;~s=1,2\}$ are arbitrary constants (determined by the specific excitation), and the remaining (frequency-dependent) constants are given by\footnote{There are sign errors in Eqs.~(8a) and (8b) of Ref.~\cite{Ebenezer08}; the RHSs of Eqs.~(8c) and (8d) of Ref.~\cite{Ebenezer08} should also be reciprocated \cite{DDEPC}.}. 
\begin{equation}
k_{zn1}=\sqrt{{\rho\omega^2\over(\lambda+2\mu)}-k^2_{rn}}, \quad k_{zn2}=\sqrt{{\rho\omega^2\over\mu}-k^2_{rn}}, \quad n=1,2,3,\ldots,N_r
\end{equation}
\begin{equation}
k_{rn1}=\sqrt{{\rho\omega^2\over(\lambda+2\mu)}-k^2_{zn}}, \quad k_{rn2}=\sqrt{{\rho\omega^2\over\mu}-k^2_{zn}}, \quad n=1,2,3,\ldots,N_z
\end{equation}
\begin{equation}
\psi_{n1}={k_{rn}\over k_{zn1}}, \quad \psi_{n2}=-{k_{zn2}\over k_{rn}}, \quad n=1,2,3,\ldots,N_r
\end{equation}
\begin{equation}
\chi_{n1}=-{k_{rn1}\over k_{zn}}, \quad \chi_{n2}={k_{zn}\over k_{rn2}}, \quad n=1,2,3,\ldots,N_z.
\end{equation}
\end{subequations} 

The components of the stress field corresponding to the displacement field (\ref{Ebensoln}) are as follows \cite{Ebenezer08}: 
\begin{eqnarray}\label{DDEsigmaRR}
\sigma^{(ENBKS)}_{rr}&=&-PK_1\lambda\sin(K_1z) \nonumber \\
&+&{\displaystyle\sum_{n=1}^{N_r}}{\displaystyle\sum_{s=1}^2}P_{ns}\left\{\left[(\lambda+2\mu)\psi_{ns}k_{rn}-\lambda k_{zns}\right]J_0(k_{rn}r)-{2\mu\over r}\psi_{ns}J_1(k_{rn}r)\right\}\sin(k_{zns}z) \nonumber \\
&+&{\displaystyle\sum_{n=1}^{N_z}}{\displaystyle\sum_{s=1}^2}Q_{ns}\left\{\left[(\lambda+2\mu)\chi_{ns}k_{rns}-\lambda k_{zn}\right]J_0(k_{rns}r)-{2\mu\over r}\chi_{ns}J_1(k_{rns}r)\right\}\sin(k_{zn}z), \nonumber \\ 
\end{eqnarray}
\begin{eqnarray}\label{DDEsigmaZZ}
\sigma^{(ENBKS)}_{zz}&=&-PK_1(\lambda+2\mu)\sin(K_1z) \nonumber \\
&+&{\displaystyle\sum_{n=1}^{N_r}}{\displaystyle\sum_{s=1}^2}P_{ns}\left[-(\lambda+2\mu)k_{zns}+\lambda\psi_{ns}k_{rn}\right]J_0(k_{rn}r)\sin(k_{zns}z) \nonumber \\
&+&{\displaystyle\sum_{n=1}^{N_z}}{\displaystyle\sum_{s=1}^2}Q_{ns}\left[-(\lambda+2\mu)k_{zn}+\lambda\chi_{ns}k_{rns}\right]J_0(k_{rns}r)\sin(k_{zn}z),
\end{eqnarray}
and 
\begin{eqnarray}\label{DDEsigmaRZ}
\sigma^{(ENBKS)}_{rz}&=&-QK_2\mu J_1(K_2r) \nonumber \\
&+&\mu{\displaystyle\sum_{n=1}^{N_r}}{\displaystyle\sum_{s=1}^2}P_{ns}\left[-k_{rn}+\psi_{ns}k_{zns}\right]J_1(k_{rn}r)\cos(k_{zns}z) \nonumber \\
&+&\mu{\displaystyle\sum_{n=1}^{N_z}}{\displaystyle\sum_{s=1}^2}Q_{ns}\left[-k_{rns}+\chi_{ns}k_{zn}\right]J_1(k_{rns}r)\cos(k_{zn}z).
\end{eqnarray}

\subsection{Application of the ENBKS Formalism to the Axisymmetric Case of BVP 2}

Since the conditions at the flat ends of the cylinder are uniform, a frequency dependence in the axial parts of the stress and displacement components can be immediately ruled out. This may also be deduced more directly from inspection of the longitudinal stress component (\ref{DDEsigmaZZ}), which must vanish at the ends of the cylinder. To enforce the vanishing of the longitudinal stress at the flat ends of the cylinder, we set the arbitrary constant $k_{zn}=(n\pi/L)$ and prescribe $P_{ns}=0,~\forall n,s$ in Eq.~(\ref{DDEsigmaZZ}) and thereby in all of the remaining ENBKS field equations. The values of the constants $\psi_{ns}$, $k_{rn}$, and $k_{zns}$ appearing in the associated summands are subsequently irrelevant. Unless the parameters are such that ${\rho\omega^2/(\lambda+2\mu)}=\left({n\pi/L}\right)^2$ (a singular case that we have ignored, see comments at the end of Sec.~\ref{GenFormDispField}), we can also choose $P=0$ in (\ref{DDEsigmaZZ}) and thereby in (\ref{DDEsigmaRR}) and (\ref{EbensolnP1}). Hence, under the condition that ${\rho\omega^2/(\lambda+2\mu)}\neq\left({n\pi/L}\right)^2$, the sub-solution (\ref{EbensolnP1}) is zero, i.e., $\left(u^{(1)}_z,u^{(1)}_r\right)=(0,0)$. The general displacement field that we seek is thus entirely contained in sub-solution (\ref{EbensolnP2}). 
 
Anticipating that we shall only require the individual harmonics of the stresses (\ref{DDEsigmaRR})-(\ref{DDEsigmaRZ}) and thereby only the corresponding individual harmonics of the displacement field, each sum over $n$ in Eqs.~(\ref{EbensolnP1},\ref{EbensolnP2},\ref{DDEsigmaRR}-\ref{DDEsigmaRZ}) collapses to a single term that depends on the prescribed value of $n$. Universally replacing the index $n$ by $k$ then yields:
\begin{eqnarray}\label{DDEconst1}
k_{rn	1} \longrightarrow k_{rk1}=\sqrt{{\rho\omega^2\over(\lambda+2\mu)}-\left({k\pi\over L}\right)^2}=\left\{\begin{array}{rrr}
             i\alpha_1 & ~~\text{Case~1} \\
             \alpha_1 & ~~\text{Case~2} \\
             i\alpha_1 & ~~\text{Case~3} \\
             \end{array} \right., 
\end{eqnarray}
\begin{eqnarray}\label{DDEconst2}
k_{rn	2} \longrightarrow k_{rk2}=\sqrt{{\rho\omega^2\over\mu}-\left({k\pi\over L}\right)^2}=\left\{\begin{array}{rrr}
             i\alpha_2 & ~~\text{Case~1} \\
             \alpha_2 & ~~\text{Case~2} \\
             \alpha_2 & ~~\text{Case~3} \\
             \end{array} \right., 
\end{eqnarray}
\begin{eqnarray}\label{DDEconst3}
\chi_{n1} \longrightarrow \chi_{k1}=-{k_{rk1}\over k_{zk}}=\left\{\begin{array}{rrr}
             \alpha_1/i\left({k\pi\over L}\right) & ~~\text{Case~1} \\
             -\alpha_1/\left({k\pi\over L}\right) & ~~\text{Case~2} \\
             \alpha_1/i\left({k\pi\over L}\right) & ~~\text{Case~3} \\
             \end{array} \right., 
\end{eqnarray}
\begin{eqnarray}\label{DDEconst4}
\chi_{n2} \longrightarrow \chi_{k2}={k_{zk}\over k_{rk2}}=\left\{\begin{array}{rrr}
             \left({k\pi\over L}\right)/i\alpha_2 & ~~\text{Case~1} \\
             \left({k\pi\over L}\right)/\alpha_2 & ~~\text{Case~2} \\
             \left({k\pi\over L}\right)/\alpha_2 & ~~\text{Case~3} \\
             \end{array} \right., 
\end{eqnarray}
where the final equalities in (\ref{DDEconst1})-(\ref{DDEconst4}) are with reference to constants $\{\alpha_1,\alpha_2\}$ as defined by Eqs.~(\ref{alphasEG}), (\ref{alphasEG5}), and (\ref{alphasEG3}), for Cases 1, 2, and 3, respectively. 

\subsubsection{Displacement Components}

For Case 1, the radial component $u^{(2)}_r$ in sub-solution (\ref{EbensolnP2}) reduces as follows:
\begin{eqnarray}\label{DDERAD2}
u^{(2)}_r&=&\left[{\displaystyle\sum_{s=1}^2}Q_{ks}\chi_{ks}J_1(k_{rks})\right]\sin(k_{zk}z) \nonumber \\ 
&=&\left[Q_{k1}{\alpha_1\over\left({k\pi\over L}\right)}i^{-1}J_1(i\alpha_1r)+Q_{k2}{\left({k\pi\over L}\right)\over\alpha_2}i^{-1}J_1(i\alpha_2r)\right]\sin\left({k\pi\over L}z\right) \nonumber \\
&=&\left[Q_{k1}{\alpha_1\over\left({k\pi\over L}\right)}I_1(\alpha_1r)+Q_{k2}{\left({k\pi\over L}\right)\over\alpha_2}I_1(\alpha_2r)\right]\sin\left({k\pi\over L}z\right) \nonumber \\
&=&\Big[\bar{A}_1{\alpha_1}I_1(\alpha_1r)+\bar{A}_2{\alpha_2}I_1(\alpha_2r)\Big]\sin\left({k\pi\over L}z\right), 
\end{eqnarray}
where relations (\ref{DDEconst1})-(\ref{DDEconst4}) have been employed in obtaining the second equality in (\ref{DDERAD2}) and the fundamental definition of the modified Bessel function of the first kind   
\begin{equation}\label{BFiden}
I_p(x)\equiv i^{-p}J_p(ix), \quad p\in\mathbb{R} 
\end{equation}
in obtaining the third equality. The last equality in (\ref{DDERAD2}) ensues upon replacement of the arbitrary constants $\left\{Q_{k1},Q_{k2}\right\}$ with a new set of arbitrary constants $\left\{\bar{A}_1,\bar{A}_2\right\}$ as follows:
\begin{eqnarray}\label{correspconstsC1}
Q_{ks} \longrightarrow\left({k\pi\over L}\right)\bar{A}_s\gamma_s=\left\{\begin{array}{lll}
             \left({k\pi\over L}\right)\bar{A}_1 & ~~\text{if}~s=1 \\
             \pm{\bar{A}_2\alpha^2_2/\left({k\pi\over L}\right)} & ~~\text{if}~s=2 \\
             \end{array} \right., 
\end{eqnarray} 
where the plus sign in (\ref{correspconstsC1}) applies to Case 1 and the minus sign to Cases 2 and 3. 

Using the same logic (and again considering Case 1), the axial component $u^{(2)}_z$ in sub-solution (\ref{EbensolnP2}) similarly reduces as follows:
\begin{eqnarray}\label{DDEAXL2}
u^{(2)}_z&=&QJ_0(K_2r)+\left[{\displaystyle\sum_{s=1}^2}Q_{ks}J_0(k_{rks}r)\right]\cos(k_{zk}z) \nonumber \\ 
&=&QJ_0(K_2r)+\Big[Q_{k1}J_0(i\alpha_1r)+Q_{k2}J_0(i\alpha_2r)\Big]\cos\left({k\pi\over L}z\right) \nonumber \\
&=&QJ_0(K_2r)+\Big[Q_{k1}I_0(\alpha_1r)+Q_{k2}I_0(\alpha_2r)\Big]\cos\left({k\pi\over L}z\right) \nonumber \\
&=&QJ_0(K_2r)+\left({k\pi\over L}\right)\Big[\bar{A}_1{\gamma_1}I_0(\alpha_1r)+\bar{A}_2{\gamma_2}I_0(\alpha_2r)\Big]\cos\left({k\pi\over L}z\right). 
\end{eqnarray}
The first term of (\ref{DDEAXL2}) is consistent with the axial component of (\ref{gensolnkeq0}) in the special $m=0$ case. Comparing (\ref{DDERAD2}) with (\ref{SolBVP2C1meq01}) and the second term of (\ref{DDEAXL2}) with (\ref{SolBVP2C1meq02}), we see that, in Case 1, the displacement components obtained from the ENBKS method are identical to the displacement components obtained from our method of solution in the special $m=0$ case. Using relations (\ref{DDEconst1}-\ref{DDEconst4},\ref{correspconstsC1}) and applying the same logic, it is straightforward to show that the equivalence holds for Cases 2 and 3 as well. Thus, in the special $m=0$ case, the general displacement field obtained from our method of solution is identical to the general axisymmetric displacement field obtained from the ENBKS method.  

\subsubsection{Stress Components}

For Case 1 and $m=0$, the radial stress component for BVP 2 obtained from our method of solution is (omitting the $\sin(\omega t)$ factor): 
\begin{eqnarray}\label{ourRRstrsscompC1meq0}
\sigma_{rr}&=&\left[\sum_{s=1}^2\bar{A}_s\left\{\left[(\lambda+2\mu)\alpha^2_s-\lambda\gamma_s\left({k\pi\over L}\right)^2\right]I_0(\alpha_sr)-{2\mu\alpha_s\over r}I_1(\alpha_sr)\right\}\right]\sin\left({k\pi\over L}z\right) \nonumber \\
&=&\left[\sum_{s=1}^2Q_{ks}\left\{\left[{(\lambda+2\mu)\alpha^2_s\over\left({k\pi\over L}\right)\gamma_s}-\lambda\left({k\pi\over L}\right)\right]I_0(\alpha_sr)-{2\mu\alpha_s\over\left({k\pi\over L}\right)\gamma_sr}I_1(\alpha_sr)\right\}\right]\sin\left({k\pi\over L}z\right) \nonumber \\
&=&\left[Q_{k1}\left\{\left[{(\lambda+2\mu)\alpha^2_1\over\left({k\pi\over L}\right)}-\lambda\left({k\pi\over L}\right)\right]I_0(\alpha_1r)-{2\mu\alpha_1\over \left({k\pi\over L}\right)r}I_1(\alpha_1r)\right\}\right. \nonumber \\ 
&&~~+\left.Q_{k2}\left\{2\mu\left({k\pi\over L}\right)I_0(\alpha_2r)-{2\mu\left({k\pi\over L}\right)\over\alpha_2 r}I_1(\alpha_2r)\right\}\right]\sin\left({k\pi\over L}z\right), 
\end{eqnarray}
where the inverse of (\ref{correspconstsC1}) (with the plus sign as pertinent to Case 1) has been employed in obtaining the second equality in (\ref{ourRRstrsscompC1meq0}) and the fact that 
\begin{eqnarray}\label{gammaequiv}
\gamma_2={1\over\left({k\pi\over L}\right)^2}\left[\left({k\pi\over L}\right)^2-{\rho\omega^2\over\mu}\right]={1\over\left({k\pi\over L}\right)^2}\left\{\begin{array}{rrr}
             \alpha^2_2 & ~~\text{Case~1} \\
             -\alpha^2_2 & ~~\text{Case~2} \\
             -\alpha^2_2 & ~~\text{Case~3} \\
             \end{array} \right.
\end{eqnarray}
in obtaining the third equality. (Note that $\gamma_1=1$ in all cases.)  
For Case 1, the ENBKS radial stress component (\ref{DDEsigmaRR}) reduces as follows:
\begin{eqnarray}\label{DDEsigmaRRreduce}
\sigma^{(ENBKS)}_{rr}&=&{\displaystyle\sum_{s=1}^2}Q_{ks}\left\{\Big[(\lambda+2\mu)\chi_{ks}k_{rks}-\lambda k_{zk}\Big]J_0(k_{rks}r)-{2\mu\over r}\chi_{ks}J_1(k_{rks}r)\right\}\sin(k_{zk}z) \nonumber \\
&=&\left[Q_{k1}\left\{\left[{(\lambda+2\mu)\alpha^2_1\over\left({k\pi\over L}\right)}-\lambda\left({k\pi\over L}\right)\right]J_0(i\alpha_1r)-{2\mu\alpha_1\over\left({k\pi\over L}\right)r}i^{-1}J_1(i\alpha_1r)\right\}\right. \nonumber \\ 
&&~~+\left.Q_{k2}\left\{2\mu\left({k\pi\over L}\right)J_0(i\alpha_2r)-{2\mu\left({k\pi\over L}\right)\over\alpha_2 r}i^{-1}J_1(i\alpha_2r)\right\}\right]\sin\left({k\pi\over L}z\right) \nonumber \\ 
&=&\left[Q_{k1}\left\{\left[{(\lambda+2\mu)\alpha^2_1\over\left({k\pi\over L}\right)}-\lambda\left({k\pi\over L}\right)\right]I_0(\alpha_1r)-{2\mu\alpha_1\over\left({k\pi\over L}\right)r}I_1(\alpha_1r)\right\}\right. \nonumber \\ 
&&~~+\left.Q_{k2}\left\{2\mu\left({k\pi\over L}\right)I_0(\alpha_2r)-{2\mu\left({k\pi\over L}\right)\over\alpha_2 r}I_1(\alpha_2r)\right\}\right]\sin\left({k\pi\over L}z\right), 
\end{eqnarray}
where relations (\ref{DDEconst1})-(\ref{DDEconst4}) have been employed in obtaining the second equality in (\ref{DDEsigmaRRreduce}) and definition (\ref{BFiden}) in obtaining the third equality. 

For Case 1 and $m=0$, the shear stress for BVP 2 obtained from our method of solution is (again omitting the $\sin(\omega t)$ factor): 
\begin{eqnarray}\label{ourRZstrsscompC1meq0}
\sigma_{rz}=\mu\left({k\pi\over L}\right)\left\{\sum_{s=1}^2\alpha_s(1+\gamma_s)\bar{A}_sI_1(\alpha_sr)\right\}\cos\left({k\pi\over L}z\right). 
\end{eqnarray}
For Case 1, the ENBKS shear stress component (\ref{DDEsigmaRZ}) reduces as follows:
\begin{eqnarray}\label{DDEsigmaRZreduce}
\sigma^{(ENBKS)}_{rz}&=&-\mu QK_2J_1(K_2r)+\mu\left\{{\displaystyle\sum_{s=1}^2}Q_{ks}\left[-k_{rks}+\chi_{ks}k_{zk}\right]J_1(k_{rks}r)\right\}\cos(k_{zk}z) \nonumber \\
&=&-\mu QK_2J_1(K_2r)+\mu\left\{Q_{k1}(2\alpha_1)i^{-1}J_1(i\alpha_1r)+Q_{k2}\left({\left({k\pi\over L}\right)^2\over\alpha_2}+\alpha_2\right)i^{-1}J_1(i\alpha_2r)\right\}\cos\left({k\pi\over L}z\right) \nonumber \\ 
&=&-\mu QK_2J_1(K_2r)+\mu\left\{{2\alpha_1}Q_{k1}I_1(\alpha_1r)+\left({\left({k\pi\over L}\right)\over\alpha_2}^2+\alpha_2\right)Q_{k2}I_1(\alpha_2r)\right\}\cos\left({k\pi\over L}z\right) \nonumber \\ 
&=&-\mu QK_2J_1(K_2r)+\mu\left\{\sum_{s=1}^2{\alpha_s(1+\gamma_s)\over\gamma_s}Q_{ks}I_1(\alpha_sr)\right\}\cos\left({k\pi\over L}z\right) \nonumber \\  
&=&-\mu QK_2J_1(K_2r)+\mu\left({k\pi\over L}\right)\left\{\sum_{s=1}^2\alpha_s(1+\gamma_s)\bar{A}_sI_1(\alpha_sr)\right\}\cos\left({k\pi\over L}z\right), 
\end{eqnarray}
where relations (\ref{DDEconst1})-(\ref{DDEconst4}) have been employed in obtaining the second equality in (\ref{DDEsigmaRZreduce}), definition (\ref{BFiden}) in obtaining the third equality, (\ref{gammaequiv}) in obtaining the fourth equality ($\gamma_1=1$ in all cases), and (\ref{correspconstsC1}) in obtaining the fifth equality. 

The first term of (\ref{DDEsigmaRZreduce}) is consistent with the shear stress component (\ref{strsssolnkeq0B}) in the special $m=0$ case. Comparing (\ref{DDEsigmaRRreduce}) with (\ref{ourRRstrsscompC1meq0}) and the second term of (\ref{DDEsigmaRZreduce}) with (\ref{ourRZstrsscompC1meq0}), we see that, in Case 1, the stress components obtained from the ENBKS method are identical to the stress components obtained from our method of solution in the special $m=0$ case. Using relations (\ref{DDEconst1}-\ref{DDEconst4},\ref{correspconstsC1},\ref{gammaequiv}) and applying the same logic, it is straightforward to show that the equivalence holds for Cases 2 and 3 as well. Thus, in the special $m=0$ case, the general stress field obtained from our method of solution is identical to the general axisymmetric stress field obtained from the ENBKS method.

\end{document}